%% file: 0-main.tex
\documentclass[preprint, dvipdfmx]{ptephy_v1}
\usepackage{braket}
\usepackage{amsthm}
\usepackage{ascmac}
\usepackage{amssymb, amsmath}
\usepackage{bm}
\usepackage{physics}
\usepackage{graphicx}
\usepackage{docmute}
\usepackage{color}
\usepackage{autobreak}
\usepackage{url}
\usepackage{times}
\usepackage[T1]{fontenc}
\usepackage[normalem]{ulem}
\usepackage[%
 setpagesize=false,%
 bookmarks=true,%
 bookmarksdepth=tocdepth,%
 bookmarksnumbered=true,%
 colorlinks=false,%
 pdftitle={},%
 pdfsubject={},%
 pdfauthor={},%
 pdfkeywords={}%
]{hyperref}
\input{newcommands}

\begin{document}
%Title of paper
% \title{Isotone Chain Protocol for $\bar{p}$-atom Spectroscopy}
% \title{Isotone Chain Study of $\bar{p}$-atom Spectroscopy;\\ Masurability of Strong Spin-orbit Contributions}
\title{Superfluid Band Theory for the Rod Phase in \\the Magnetized Inner Crust Matter:\\
Entrainment, Spin-orbit Coupling, and Spin-triplet Pairing}

% repeat the \author .. \affiliation  etc. as needed
% \email, \thanks, \homepage, \altaffiliation all apply to the current
% author. Explanatory text should go in the []'s, actual e-mail
% address or url should go in the {}'s for \email and \homepage.
% Please use the appropriate macro foreach each type of information

% \affiliation command applies to all authors since the last
% \affiliation command. The \affiliation command should follow the
% other information
% \affiliation can be followed by \email, \homepage, \thanks as well.

\author{Kenta Yoshimura}
%\email[]{yoshimura.k.ak@m.titech.ac.jp}
% \affil{Department of Physics, School of Science, Tokyo Institute of Technology, Tokyo 152-8550, Japan}

\author{Kazuyuki Sekizawa}
%\email[]{sekizawa@phys.sci.isct.ac.jp}
\affil{Department of Physics, School of Science, Institute of Science Tokyo, Tokyo 152-8550, Japan}

\date{\today}

\input{01-abstract}

\maketitle
\input{1-Introduction}
\input{2-Formalism}

\input{3-setting}

\input{4-result}

\input{5-summary}

\input{6-acknowledgement}

% \appendix 

% \input{A-appendix}

% \newpage
\bibliographystyle{ptephy}
\bibliography{triplet}

\end{document}

%% file: newcommands.tex
\newcommand{\up}{{\!\uparrow}}
\newcommand{\down}{{\!\downarrow}}

\newcommand{\mev}{{\mathrm{MeV}}}

%% file: 01-abstract.tex
\begin{abstract}
The inner crust of neutron stars hosts a rich variety of nuclear phenomena and provides a unique opportunity for exploring microscopic nuclear properties relevant to diverse astrophysical observations.
In particular, magnetars, which possess extremely strong magnetic-field, have attracted increasing interest in connection with nuclear spin dynamics and unconventional pairing correlations.
This work is dedicated to develop a comprehensive theoretical framework to describe structures and properties of two-dimensional crustal matter (i.e., rod phase) in the inner crust of neutron stars, incorporating band-structure effects, neutron spin-triplet pairing, and strong magnetic-fields on an equal footing.
Nuclear interactions and correlations are treated within the framework of nuclear density functional theory combined with a local description of pairing correlations, known as the superfluid local density approximation.
Within this framework, the definition of the anomalous density is generalized to include spin-triplet components, and a spin-triplet pairing energy density functional is consistently incorporated.
The main results of this study can be summarized as follows.
In the first place, magnetic-fields of the order of $10^{16}\,$G are found to substantially enhance the neutron effective mass by a factor of approximately $1.5$, indicating a significant modification of entrainment properties in strongly magnetized crustal matter.
In the second place, while the overall behavior of pairing phase transitions is qualitatively similar to that observed in one-dimensional systems studied previously, the present two-dimensional calculations reveal a nontrivial role of the spin-orbit interaction in inducing spin-polarization under magnetic-fields.
In the third place, concerning the spin-triplet superfluidity, the rank-0 component is shown to emerge as a consequence of magnetic-field-induced spin-polarization, irrespective of the presence of spin-triplet pairing interactions, whereas the rank-2 component appears only when the corresponding interaction channel is included.
These findings provide new microscopic insights into superfluid and magnetic phenomena in the inner crust of neutron stars and offer a theoretical basis for future studies of magnetars and related astrophysical processes.
\end{abstract}

%% file: 1-Introduction.tex
\section{INTRODUCTION}\label{sec:intro}
Neutron stars are compact objects which serve as natural laboratories for probing the nuclear and hadronic properties through the verification with the astronomical observations~\cite{ozel2016, lattimer2021}.
The inner crust region in a neutron star is composed of the Coulomb lattice of nuclei coexisting with free neutron gas~\cite{ravenhall1983, hashimoto1984, chamel2008, kobyakov2013, caplan2017}, where neutron superfluidity plays an indispensable role~\cite{lombardo2001, dean2003, sedrakian2019, chamel2024}.
Superfluid properties in the neutron-star crust are believed to have impacts on various astrophysical phenomena associated with neutron stars, such as pulsar glitches and neutron-star cooling~\cite{link1999, yakovlev2004, peng2006, shternin2011, zhou2022}. 
However, their detailed behavior can be altered radically by effects of the crystalline structures of nuclei including nuclear pasta phases, making a comprehensive analysis extremely challenging.
For example, band-theoretical calculations applied to the inner-crust matter have suggested that the \textit{effective mass} of free neutrons may be enhanced by factors of several up to around ten due to crystalline-structure effects~\cite{carter2005, chamel2005, chamel2012, chamel2017}. 
This phenomenon, referred to as the \textit{entrainment} effect, continues to be a subject of intense controversy,
which involves consistency with pulsar glitch interpretations~\cite{andersson2012, chamel2013, haskell2015}.
Subsequently, calculations incorporating self-consistency and superfluidity have been performed~\cite{kashiwaba2019, watanabe2017, minami2022}, and a variety of methodologies have been proposed, including approaches in which the effective mass is extracted from the dynamical response through time evolutions~\cite{sekizawa2022, yoshimura2024a}, as well as methods to determine the superfluid fraction from its static properties~\cite{almirante2024, almirante2024a, almirante2025, almirante2025a}. 
Within these researches regarding the neutron effective mass or superfluid fraction, it was often the case that only the $^1\text{S}_0$ component was taken into considerations.
However, in the context of astronomical simulations, the impact of spin-triplet superfluidity, such as the $^3\text{P}_2$ component, has been actively investigated~\cite{chau1992, schwenk2004, page2004, page2009, page2011, shternin2011, haskell2015, potekhin2015, masuda2016, masuda2020, leinson2020, masaki2020, andersson2021, kobayashi2023, marmorini2024, granados2025, nam2025, hattori2025, sedrakian2025}, and these studies are pushing their physical interpretation to the next stage.
Although spin-triplet components are generally expected to emerge only in the neutron-star core region~\cite{sedrakian2019, wei2020}, recent studies have begun to suggest their existence in the crustal matter or even in finite nuclei.
For example, in Ref.~\cite{hinohara2024}, it has been pointed out that, owing to the contribution of the spin-orbit interaction, $^3\text{P}_1$ superfluid components emerge in several open-shell nuclei.
Another example is Ref.~\cite{tajima2023} where they revealed that spin-polarized neutron matter may produce the $^3\text{P}_0$ pairing component.
There is a kind of neutron stars known as \textit{magnetars}~\cite{makishima2014, turolla2015, kaspi2017}, which possess surface magnetic-fields exceeding $10^{14}\,$G and possibly reaching up to $10^{17}$ or $10^{18}$\,G~\cite{potekhin1996a, broderick2000a, bonanno2003, naso2008, frieben2012}.
In such matter with enormous magnetic-fields, the nuclear compositions and superfluid phase structures may be drastically converted~\cite{muzikar1980, broderick2000a, penaarteaga2011, chamel2012a, basilico2015, stein2016, stein2016a, yasui2019, parmar2023, sekizawa2023, jiang2024, basilico2025}, which should be thoroughly investigated based on microscopic theories.

In our previous studies, under the title of ``superfluid band theory,'' we have pursued an approach in which superfluidity and band-structure effects are incorporated on the same footing into the density functional theory (DFT)~\cite{hohenberg1964, kohn1965, nakatsukasa2016, colo2020} for the superfluid systems~\cite{yoshimura2024a}. 
There has been a wealth of literature where the DFT calculations successfully elucidated structures and properties of the inner crust matter, including pasta phases~\cite{magierski2002a, gogelein2008a, newton2009a, pais2012a, grill2014a, schuetrumpf2015, fattoyev2017, schuetrumpf2019, pecak2024}.
For the superfluid extension of the nuclear DFT, we employ a local treatment of pairing interactions, known as superfluid local density approximation (SLDA)~\cite{bulgac2002, bulgac2002a, jin2021}.
% and accurately captured not only the nuclear pairing nature, but also nuclear static and dynamical properties~\cite{bertsch2009, bulgac2012, dobaczewski2012, bulgac2013, goriely2013, bulgac2016, zhao2016}.
By extending the framework into systems at finite temperatures and under finite magnetic-field, we have further advanced the development of a tool to comprehensively describe the structures and properties of neutron star matter under extreme conditions, such as supernova matter and magnetars~\cite{yoshimura2025}.
% In the present study, we further extend this framework to two-dimensional nuclear matter in the rod phase. 
% We focus on the rod phase as a representative geometry of nuclear pasta to investigate how its low-dimensional crystalline structure, combined with spin-orbit interactions, influences the emergence of the neutron superfluidity and spin-polarization.
% Additionally, we investigate how these properties will be modified by the effect of the spin-triplet pairing interactions, as well as magnetic-fields.
% For this purpose, it is necessary to extend the definition of the anomalous density, which has been associated with the $^1S_0$ pairing order parameter within the normal superfluid DFT, and to carry out a theoretical investigation of how spin-triplet components can exist in such a framework.
% Additionally, we can consider the interactions corresponding to the spin-triplet pairings.
% Performing calculations with and without these new contributions, we investigate how the order parameters of the spin-triplet pairing appear.
In the present study, we extend our superfluid band theory to two-dimensional nuclear matter in the rod phase, which serves as a representative geometry of nuclear pasta.
Our aim is to clarify how low-dimensional crystalline structures, in combination with spin-orbit interactions and strong magnetic-fields, influence neutron superfluidity, spin polarization, and pairing phases.
To disentangle the roles of band-structure effects, magnetic-fields, and pairing correlations, we adopt a stepwise approach: we firstly analyze the entrainment effect within the Hartree-Fock (HF) framework (i.e., without pairing correlations), and subsequently incorporate superfluidity to investigate the interplay between spin-polarization and spin-singlet pairing, under the effect of the spin-orbit interaction.
We then extend the superfluid density functional framework to include spin-triplet pairing interactions, and systematically examine the emergence of spin-triplet superfluid components by performing calculations with and without these interactions under various magnetic-field strengths.
Through this unified and self-consistent framework, this work aims to provide microscopic insights into superfluid and magnetic phenomena in neutron-star crustal matter under extreme conditions.

The article is organized as follows.
In Section 2, the theoretical framework of this study is briefly explained, including the spin-triplet pairing.
Section 3 indicates the computational settings used in the present work.
The results of the calculations are shown and discussed in Section 4.
Section 5 summarizes the work and suggests future prospects.

%% file: 2-Formalism.tex
\section{FORMALISM}
In this section, we present the theoretical framework employed in this study, starting from the generic formulations of the Skyrme HF and Hartree-Fock-Bogoliubov (HFB) approaches.
The formulation of the superfluid band theory has been developed in our previous work~\cite{yoshimura2024a}.
For the spin-triplet pairing terms, detailed derivations can be found in Refs.~\cite{dobaczewski1984, duguet2014, hinohara2024}, although the notation used therein differs slightly from the present formulation.
The extension to finite magnetic fields is based on the general formalisms discussed in Refs.~\cite{chamel2012a, stein2016}, and their implementation within our framework has been established in Ref.~\cite{yoshimura2025}.

\subsection{Skyrme-type energy density functional and Skyrme Kohn-Sham equations}
The total energy per nucleon in the neutron star matter is, in general, written as
\begin{equation}
    E/A = \frac{1}{A}\int\dd\bm{r}\, \qty[\mathcal{E}_\text{nucl}(\bm{r}) + \mathcal{E}_\text{elec}(\bm{r})],
\end{equation}
where the nuclear part of the energy density is the sum of kinetic, interaction, Coulomb, and (for superfluid systems) pairing contributions,
\begin{equation}
    \mathcal{E}_\text{nucl}(\bm{r}) = \mathcal{E}_\text{kin}(\bm{r}) + \mathcal{E}_\text{int}(\bm{r}) + \mathcal{E}_\text{Coul}(\bm{r}) +\mathcal{E}_\text{pair}(\bm{r}).
\end{equation}
The kinetic and Coulomb parts are given by
\begin{align}
    \mathcal{E}_\text{kin}(\bm{r}) &= \sum_{q=n,p}\frac{\hbar^2}{2m_q}\tau_q(\bm{r}),\\
    \mathcal{E}_\text{Coul}(\bm{r}) &= \frac{e^2}{2}\int\dd\bm{r}^\prime\, \frac{\rho_p(\bm{r})\rho_p(\bm{r}^\prime)}{\abs{\bm{r}-\bm{r}^\prime}} + \frac{3e^2}{4}\qty(\frac{3}{\pi})^{1/3}\rho_p^{4/3}(\bm{r}),
\end{align}
where the exchange term of the Coulomb energy is evaluated using the Slater approximation.
For the interaction part, we adopt the Skyrme-type EDF which possesses the time-even and time-odd components, provided as
\begin{equation}
    \begin{aligned}
        \mathcal{E}_\text{int}(\bm{r}) &= \mathcal{E}_\text{even}(\bm{r}) + \mathcal{E}_\text{odd}(\bm{r}),\\
        \mathcal{E}_\text{even}(\bm{r}) &= \sum_{t=0,1}\big[C^\rho_t \rho_t^2 + C^{\alpha}_t\rho_t^2\rho_0^\alpha + C^{\Delta\rho}_t \rho_t\Delta\rho_t+ C^\tau_t\rho_t\tau_t  + C^{\nabla J}\rho_t\grad\cdot\bm{J}_t - C^T_t\sum_{\mu,\nu=x}^z (J_{t,\mu\nu})^2 \big],\\
        \mathcal{E}_\text{odd}(\bm{r}) &= \sum_{t=0,1}\big[C^s_t\bm{s}_t^2 + C^{s\alpha}_t \rho^{\alpha}_0\bm{s}_t^2 + C^T_t\bm{s}_t\cdot\bm{T}_t + C^{\Delta s}_t\bm{s}_t\cdot\Delta\bm{s}_t  -C^\tau_t\bm{j}_t^2 + C^{\nabla J}\bm{s}_t\cdot(\grad\times\bm{j}_t).
    \end{aligned}
\end{equation}
The coupling constants $C_t^\text{X}$, where X stands for $\rho$, $\tau$, etc., are expressed in terms of the Skyrme force parameters, as summarized in Ref.~\cite{lesinski2007}.
Since the $\bm{s}_t\cdot\Delta\bm{s}_t$ term is known to induce spin instabilities~\cite{hellemans2012, sekizawa2013}, we set $C_t^{\Delta s}=0$ throughout this work.
The local densities $\rho$, $\tau$, $\bm{j}$, $\bm{s}$, $\bm{T}$, and $J_{\mu\nu}$ denote the particle density, kinetic density, current density, spin density, spin-kinetic density, and spin-current tensor density, respectively.
They are constructed from the single-particle wave functions as
\begin{eqnarray}
    \rho(\bm{r})&=& \sum_{k\sigma} \abs{\psi_{k\sigma}(\bm{r})}^2,\\
    \tau(\bm{r})&=& \sum_{k\sigma} \abs{\grad \psi_{k\sigma}(\bm{r})}^2,\\
    \bm{j}(\bm{r}) &=& \sum_{k\sigma} \Im[\psi^*_{k\sigma}(\bm{r})\grad \psi_{k\sigma}(\bm{r})],\\
    \bm{s}(\bm{r}) &=& \sum_{k,s,s^\prime} \bm{\sigma}_{ss^\prime} \psi^*_{ks}(\bm{r})\psi_{ks^\prime}(\bm{r}),\\
    \bm{T}(\bm{r}) &=& \sum_{k,s,s^\prime} \bm{\sigma}_{ss^\prime} \grad \psi^*_{ks}(\bm{r})\cdot \grad\psi_{ks^\prime}(\bm{r}),\\
    J_{\mu\nu}(\bm{r}) &=& \sum_{k,s,s^\prime}\frac{1}{2i}\qty[\sigma_{\nu}]_{ss^\prime}\big[\psi^*_{ks^\prime}(\bm{r})\qty(\nabla_\mu \psi_{ks}(\bm{r}))- \psi_{ks}(\bm{r})\qty(\nabla_\mu \psi_{ks^\prime}(\bm{r}))\big],\label{eq:spin-current-hf}
\end{eqnarray}
whrere $\bm{\sigma}=(\sigma_x,\sigma_y,\sigma_z)$ denotes the Pauli matrices.
The spin-current tensor can be decomposed into its scalar (trace), vector, and traceless symmetric components as
\begin{equation}
    \begin{aligned}
        J_t(\bm{r}) &= \sum_{\mu=x,y,z} J_{t,\mu\mu}(\bm{r})\\
        \bm{J}_{t,\mu}(\bm{r}) &= \sum_{\nu,\lambda=x,y,z}\epsilon_{\mu\nu\lambda} J_{t,\nu\lambda}(\bm{r})\\
        \underline{\mathsf{J}}_{t,\mu\nu}(\bm{r}) &= \frac{1}{2}J_{t,\mu\nu}(\bm{r}) + \frac{1}{2}J_{t,\nu\mu}(\bm{r}) - \frac{1}{3}J_{t}(\bm{r})\delta_{\mu\nu}.\label{eq:current-decomp}
    \end{aligned}
\end{equation}
The ground state is obtained by minimizing the energy functional under the orthonormality constraint~\cite{maruhn2014, jin2021},
\begin{equation}
    \delta \qty(E - \sum_i \epsilon_i \braket{\phi_i}{\phi_i}) = 0,
\end{equation}
leading to the spin-dependent Kohn-Sham equation:
\begin{equation}
    \hat{h}_q\mqty(\psi^{(q)}_i(\bm{r}\up) \\ \psi^{(q)}_i(\bm{r}\down)) = \epsilon_i \mqty(\psi^{(q)}_i(\bm{r}\up) \\ \psi^{(q)}_i(\bm{r}\down)),\label{eq:SkyrmeHF}
\end{equation}
where $q=n,p$ denotes the nucleon species.
The single-particle Hamiltonian takes the general form
\begin{equation}
    \begin{aligned}
        \hat{h}(\bm{r}) &= -\grad\cdot\qty(M(\bm{r}) + \bm{\sigma}\cdot\bm{\Lambda}(\bm{r}))\grad + U(\bm{r}) + \bm{\sigma}\cdot\bm{\Sigma}(\bm{r})\\
        &\hspace{10mm} + \frac{1}{2i}\qty[\bm{W}(\bm{r})\cdot(\grad\times\bm{\sigma}) + \grad\cdot(\bm{\sigma}\times \bm{W}(\bm{r}))]+ \frac{1}{2i}\qty[\grad\cdot \bm{I}(\bm{r}) + \bm{I}(\bm{r})\cdot\grad]\\
        &\hspace{10mm} + \frac{1}{2i}\qty[\grad\cdot\qty(\mathsf{B}\cdot\bm{\sigma}) + \qty(\mathsf{B}\cdot\bm{\sigma}) \cdot\grad],
    \end{aligned}
\end{equation}
where the mean-field potentials can be obtained from functional derivatives,
\begin{align*}
    \frac{\delta E}{\delta \rho} &= U(\bm{r}),& \frac{\delta E}{\delta \tau}&=M(\bm{r}),\\
    \frac{\delta E}{\delta \bm{j}} &= \bm{I}(\bm{r}),&\frac{\delta E}{\delta \bm{s}} &= \bm{\Sigma}(\bm{r}),\\
    \frac{\delta E}{\delta \bm{T}} &= \bm{\Lambda}(\bm{r}),&
    \frac{\delta E}{\delta \bm{J}} &= \bm{W}(\bm{r}),\\
    \frac{\delta E}{\delta J_{\mu\nu}} &= \mathsf{B}_{\mu\nu}(\bm{r}).
\end{align*}
Here we have omitted the isospin index $q$ for simplicity.
These potentials are explicit functionals of the local densities, and the ground state is obtained self-consistently by iteratively solving the Kohn-Sham equations.

\subsection{Band theory for inner crust matter}
The central idea of band theory is encapsulated in the Bloch's theorem,
\begin{equation}
    \psi(\bm{r}) = \tilde{\psi}(\bm{r})e^{i\bm{k}\bm{r}},\quad \tilde{\psi}(\bm{r}+\bm{R}) = \tilde{\psi}(\bm{r}),
\end{equation}
where $\bm{k}$ is the Bloch wave vector and $\bm{R}$ is a lattice translation vector.
Substituting the Bloch form into the Kohn-Sham equation~\eqref{eq:SkyrmeHF}, we obtain
\begin{equation}
    \hat{h}_{q,\bm{k}}\mqty(\tilde{\psi}^{(q)}_{i\bm{k}}(\bm{r}\up) \\ \tilde{\psi}^{(q)}_{i\bm{k}}(\bm{r}\down)) = \epsilon_{i\bm{k}} \mqty(\tilde{\psi}^{(q)}_{i\bm{k}}(\bm{r}\up) \\ \tilde{\psi}^{(q)}_{i\bm{k}}(\bm{r}\down)),
\end{equation}
where $\hat{h}_{q,\bm{k}}$ can be obtained from $\hat{h}_q$ by replacing derivatives as
\begin{equation}
    \grad \to  \grad + i\bm{k}.\label{eq:derivative-shift}
\end{equation}
In the present study, we focus on the two-dimensional crystalline configuration corresponding to the rod (or ``spaghetti'') phase in the neutron-star inner crust.
The system is assumed to be uniform along the $z$ direction, and hence the cell-periodic functions $\tilde{\psi}_{i\bm{k}}(\bm{r})$ depend only on the transverse coordinates $(x,y)$.
As a consequence, spatial derivatives with respect to $z$ acting on $\tilde{\psi}_{i\bm{k}}$ vanish, while the longitudinal Bloch wave number $k_\parallel$ associated with the $z$ direction is retained.
In this formulation, the normalization is imposed on the full Bloch wave functions $\psi_{i\bm{k}}(\bm{r})$, rather than on the cell-periodic part $\tilde{\psi}_{i\bm{k}}(\bm{r})$.
Specifically, the normalization is carried out over a supercell consisting of $N_{k_x}\times N_{k_y}$ unit cells in the $x$ and $y$ directions, respectively.
Under this convention, the normalization condition for the cell-periodic function
$\tilde{\psi}_{i\bm{k}}$ takes the form
\begin{equation}
    \int \dd x\,\dd y\, \tilde{\psi}_{i\bm{k}}(x,y)
    = aN_{k_x} bN_{k_y}.
\end{equation}
where $a$ and $b$ denote the lattice constants along the $x$ and $y$ directions, respectively.
In the present calculations, we restrict ourselves to a square lattice, i.e., $a=b$.
With this choice of normalization, local densities are defined as spatial averages over the supercell.
Accordingly, explicit factors of $(aN_{k_x})(bN_{k_y})$ appear in the expressions for the densities, reflecting the normalization volume of the Bloch states.
Derivatives acting on the Bloch wave functions are evaluated using the shifted operator~\eqref{eq:derivative-shift}.
For example, the particle density, kinetic energy density, and current density
are given by
\begin{align}
    \rho_q(x,y) &= \sum_{i\bm{k}\sigma} \frac{dk_\parallel}{2\pi(aN_{k_x})(bN_{k_y})}\abs{\tilde{\psi}^{(q)}_{i\bm{k}}(\bm{r}\sigma)}^2\\
    \tau_q(x,y) &= \sum_{i\bm{k}\sigma} \frac{dk_\parallel}{2\pi(aN_{k_x})(bN_{k_y})}\abs{\grad' \tilde{\psi}^{(q)}_{i\bm{k}}(\bm{r}\sigma)}^2\\
    \bm{j}_q(x,y) &= \sum_{i\bm{k}\sigma} \frac{dk_\parallel}{2\pi(aN_{k_x})(bN_{k_y})}\Im\qty[\tilde{\psi}^{(q)*}_{i\bm{k}}(\bm{r}\sigma)\grad' \tilde{\psi}^{(q)}_{i\bm{k}}(\bm{r}\sigma)],
\end{align}
where the shifted derivative operator in the present quasi-two-dimensional setup is given by
\begin{equation}
    \grad' = (\nabla_x + ik_x, \nabla_y + ik_y, ik_\parallel).
\end{equation}
The remaining local densities are constructed in an analogous manner.

The band-structure effects on dripped neutrons in the inner crust are quantified in terms of the macroscopic effective mass~\cite{carter2005, kashiwaba2019}, defined through
\begin{equation}
    m_n n^f_n = m^\star_{n,\mu} n^c_{n,\mu},\label{eq:effectivemass}
\end{equation}
where $n_n^{f}$ and $n_{n,\mu}^{c}$ denote the free neutron density and the conduction neutron density, respectively.
The free neutron density counts neutrons that are not bound in the nuclear mean field and is evaluated as
\begin{equation}
    n^f_n = \frac{1}{(aN_{k_x}) (bN_{k_y})}\sum_{i\bm{k}}\frac{\dd k_\parallel}{2\pi} \theta(\mu_n - \epsilon^{(n)}_{i\bm{k}})\theta(\epsilon^{(n)}_{i\bm{k}} - U^{(\text{max})}_n),
\end{equation}
where $\mu_n$ is the neutron chemical potential and $U_n^{(\mathrm{max})}$ is the maximum value of the neutron single-particle potential within the unit cell.
This definition corresponds to the number of neutrons that would be mobile in the absence of band-structure effects.
The conduction neutron density is defined in terms of the mobility tensor,
\begin{equation}
    n^c_{n,\mu} = m_n \mathcal{K}^{\mu\mu},
\end{equation}
where the mobility coefficient is calculated from the curvature of the energy bands,
\begin{equation}
    \mathcal{K}^{\mu\nu} = \frac{1}{(aN_{k_x})(bN_{k_y})}\sum_{i\bm{k}} \frac{\dd k_\parallel}{2\pi} \frac{\partial^2 \epsilon^{(n)}_{i\bm{k}}}{\partial k_\mu \partial k_\nu}\theta(\mu_n - \epsilon^{(n)}_{i\bm{k}}).\label{eq:mobilitycoefficient}
\end{equation}
The curvature of the band dispersion reflects the mobility of particles in the crystal, and the conduction density therefore measures the fraction of dripped neutrons that can move effectively in the presence of band-structure effects.
Although we previously proposed a time-dependent approach to extract the effective mass directly~\cite{yoshimura2024a}, we employ the above standard band-theoretical definition in the present work, because our primary interest here is the modification of single-particle band structures induced by magnetic-fields and spin-orbit coupling.

\subsection{Hartree-Fock-Bogoliubov theory}
In the HFB theory~\cite{ring2004}, quasiparticle creation and annihilation operators $\hat{\beta}^\dag$ and $\hat{\beta}$ are defined through the Bogoliubov transformation of the particle operators $\hat{a}^\dag$ and $\hat{a}$ as
\begin{equation}
    \mqty(\hat{\bm{\beta}} \\ \hat{\bm{\beta}}^\dag) = \mqty(U^\dag & V^\dag \\ V^{\mathsf{T}} & U^{\mathsf{T}})\mqty(\bm{\hat{a}} \\ \bm{\hat{a}}^\dag).
\end{equation}
The HFB ground state $\ket{\Phi}$ is defined as the quasiparticle vacuum,
\begin{equation}
    \hat{\beta}_\mu \ket{\Phi} = 0\quad \text{for}\,^\forall\mu.
\end{equation}
The coordinate-space representation is constructed using the single-particle wave functions and field operators
\begin{equation}
    \phi_i(\bm{r}\sigma) = \bra{\bm{r}\sigma}\hat{a}^\dag_i \ket{0},\quad
    \hat{\psi}(\bm{r}\sigma) = \sum_i \phi_i(\bm{r}\sigma) \hat{a}_i.
\end{equation}
Within this framework, one can define not only the normal (particle) density but also the anomalous density,
\begin{equation}
    \begin{aligned}
        \rho(\bm{r}\sigma,\bm{r}'\sigma') &= \bra{\Phi} \psi^\dag(\bm{r}'\sigma')\psi(\bm{r}\sigma) \ket{\Phi},\\
        \kappa(\bm{r}\sigma,\bm{r}'\sigma') &= \bra{\Phi} \psi(\bm{r}'\sigma')\psi(\bm{r}\sigma) \ket{\Phi}.
    \end{aligned}
\end{equation}
These densities can be expressed in terms of the quasiparticle wave functions as
\begin{equation}
        \rho(\bm{r}\sigma,\bm{r}'\sigma') = \sum_\mu v_\mu(\bm{r}'\sigma') v^*_\mu (\bm{r}\sigma),\quad
        \kappa(\bm{r}\sigma,\bm{r}'\sigma') = \sum_\mu u_\mu(\bm{r}'\sigma')v^*_\mu(\bm{r}\sigma), 
\end{equation}
with quasiparticle wave functions
\begin{equation}
        u_{\mu}(\bm{r}\sigma) = \sum_i U_{i\mu}\phi_i(\bm{r}\sigma),\quad 
        v_\mu(\bm{r}\sigma) = \sum_i V_{i\mu}\phi_i^*(\bm{r}\sigma).
\end{equation}

Owing to the fermionic anti-commutation relations, the anomalous density satisfies the antisymmetric condition
\begin{equation}
    \kappa(\bm{r}\sigma,\bm{r}'\sigma') = -\kappa(\bm{r}'\sigma',\bm{r}\sigma).
\end{equation}
Within the superfluid local density approximation (SLDA), this property reduces to a purely spin-antisymmetric relation,
\begin{equation}
    \kappa_{\sigma\sigma'}(\bm{r}) = -\kappa_{\sigma'\sigma}(\bm{r}).
\end{equation}
which allows the local anomalous density to be written in the form
\begin{equation}
    \kappa_{\sigma\sigma'}(\bm{r})
    = \qty(i\sigma_y\,\kappa(\bm{r}))_{\sigma\sigma'} .
\end{equation}
Here, $\kappa_{\sigma\sigma'}$ denotes the anomalous density matrix in the spin space, while $\kappa(\bm{r})$ represents the pairing order parameter expressed as a
$2\times2$ spin matrix.
The pairing order parameter can be decomposed into spin-singlet and spin-triplet components~\cite{sigrist1991, sigrist2005},
\begin{equation}
    \kappa(\bm{r}) = \kappa^{(s)}(\bm{r}) + \bm{\sigma}\cdot \bm{\kappa}^{(t)}(\bm{r}),
\end{equation}
where the superscripts $(s)$ and $(t)$ denote singlet and triplet contributions, respectively.
These components are extracted as
\begin{align}
    \kappa^{(s)}(\bm{r}) = \frac{1}{2}\Tr{-i\sigma_y\kappa_{\sigma\sigma'}(\bm{r})},\quad
    \kappa^{(t)}_\mu(\bm{r}) = \frac{1}{2}\Tr{-i\sigma_y\sigma_\mu \kappa_{\sigma\sigma'}(\bm{r})},
\end{align}
where the trace is taken over spin indices.
Assuming that only isovector ($T=1$) pairing channels ($nn$ and $pp$) are relevant, the allowed pairing modes are restricted to $s$-wave spin-singlet and $p$-wave spin-triplet components.
The singlet anomalous density reduces to the familiar expression,
\begin{equation}
    \kappa^{(s)}(\bm{r}) = \frac{1}{2}\qty(\kappa_{\up\down}(\bm{r}) - \kappa_{\down\up}(\bm{r})) = \kappa_{\up\down}(\bm{r}).
\end{equation}
The spin-triplet pairing is characterized by the pairing spin-current tensor~\cite{hinohara2024},
\begin{equation}
    \tilde{J}_{\mu\nu}(\bm{r}) = -\frac{i}{2}\qty(\partial_\mu -\partial'_\mu)\kappa^{(t)}_\nu(\bm{r},\bm{r}')\eval_{\bm{r}=\bm{r}'}.
\end{equation}
The pairing energy is written as
\begin{align}
    E_\text{pair}^{S=0} &= \int\dd\bm{r}\, \sum_t g_{\text{eff},t}(\bm{r})\abs{\kappa^{(s)}_t(\bm{r})}^2,\\
    E_\text{pair}^{S=1} &= \int \dd\bm{r}\, \sum_t \tilde{C}^J_t \abs{\tilde{J}_t(\bm{r})}^2,
\end{align}
where $g_{\mathrm{eff}}$ denotes the effective pairing strength~\cite{jin2021}.
We note that $\tilde{C}^J_t$ is different from the coefficients included in the standard Skyrme EDF $C^J_t$. 
Using the same decomposition as in Eq.~\eqref{eq:current-decomp}, the spin-triplet pairing energy can be rewritten as
\begin{equation}
    E^{S=1}_\text{pair} = \sum_{M=0}^2 \int\dd\bm{r}\,\sum_t \tilde{C}^{JM}_t\abs{\tilde{J}^M_t(\bm{r})}^2,
\end{equation}
where $\tilde{J}^{M=0}_t$, $\tilde{J}^{M=1}_t$, and $\tilde{J}^{M=2}_t$ correspond to the scalar, vector, and traceless symmetric components of the pairing spin-current, respectively.
Although these coupling constants are constrained as $\tilde{C}^J_t = 3\tilde{C}^{J0}_t = 2\tilde{C}^{J1}_t = \tilde{C}^{J2}_t$,
we treat them as independent parameters in the present study to allow for general spin-triplet pairing interactions.
% Although these associated coupling constants should satisfy $\tilde{C}^J_t = 3\tilde{C}^{J0}_t = 2\tilde{C}^{J1}_t = \tilde{C}^{J2}_t$ in principle, we can assume that they take independent values considering \textit{e.g.} the pairing spin interactions.
The specific values of these constants used in this study will be provided in the later section.
Under the spherical symmetric systems, the decomposed spin-current components $\tilde{J}^M_t$ are connected with the order parameters of the $^3\text{P}_M$ pairing.
However, the present rod phase does not have such symmetry and it is not clear whether these pairing spin-current components and order parameters can be exactly related.
This is why we call these contributions simply rank-0, rank-1, and rank-2 components, respectively.
Nevertheless, we note that it is possible that they could be approximately connected even in the non-spherical rod phases.

The HFB equations take the standard matrix form
\begin{equation}
    \mqty(\hat{h}-\mu & \Delta \\ -\Delta^* & \hat{h}^*+ \mu)\mqty(u_{\mu}(\bm{r}) \\ v_{\mu}(\bm{r})) = E_\mu \mqty(u_{\mu}(\bm{r}) \\ v_{\mu}(\bm{r})),
\end{equation}
where $\hat{h}$ and $\Delta$ are $2\times2$ matrices in the spin space, as well as wave functions $u$ and $v$ are spin vectors.
The spin components of the pairing field $\Delta$ are also reduced to the spin-singlet and triplet parts, which is
\begin{equation}
    \Delta_{\sigma\sigma'}(\bm{r}) = (i\sigma_y)_{\sigma\sigma'}\Delta^{(s)}(\bm{r}) + (i\sigma_y\bm{\sigma})_{\sigma\sigma'} \cdot \bm{\Delta}^{(t)}(\bm{r}),
\end{equation}
where
\begin{align}
    \Delta^{(s)}(\bm{r}) &= g_\text{eff}(\bm{r})\kappa^{(s)}(\bm{r}),\\
    (i\sigma_y\bm{\sigma})\cdot \bm{\Delta^{(t)}} &= -\frac{i}{2}\qty[\grad\cdot\qty[\tilde{\mathsf{B}}\cdot(i\sigma_y\bm{\sigma})] + \qty[\tilde{\mathsf{B}}\cdot(i\sigma_y\bm{\sigma})] \cdot\grad],
\end{align}
where the pairing mean field $\tilde{\mathsf{B}}$ is a spacial tensor and obtained from the functional derivative of the spin-triplet pairing energy,
\begin{equation}
    \tilde{\mathsf{B}}_{\mu\nu}(\bm{r}) = 2\tilde{C}^{J0}\tilde{J}(\bm{r})\delta_{\mu\nu} + 2\tilde{C}^{J1}\sum_{\lambda}\epsilon_{\lambda\mu\nu} \tilde{\bm{J}}_{\lambda}(\bm{r}) + 2\tilde{C}^{J2}\tilde{\underline{\mathsf{J}}}_{\mu\nu}(\bm{r}).
\end{equation}
When the band-structure effects are taken into account, the HFB Hamiltonian turns into
\begin{equation}
    \mqty(\hat{h}_{\bm{k}}-\mu & \Delta_{\bm{k}} \\ -\Delta_{-\bm{k}}^* & \hat{h}^*_{-\bm{k}} + \mu)\mqty(u_{\mu\bm{k}}(\bm{r}) \\ v_{\mu\bm{k}}(\bm{r})) = E_{\mu\bm{k}} \mqty(u_{\mu\bm{k}}(\bm{r}) \\ v_{\mu\bm{k}}(\bm{r})),
\end{equation}
where the indices $\bm{k}$ indicate the same derivative shift \eqref{eq:derivative-shift}.
It should be taken care that terms belonging to $v_{\mu\bm{k}}$ basis yield $-\bm{k}$ shift instead of $\bm{k}$.

Finally, following our previous work, we quantify the spin-triplet condensations by the integrated quantities
\begin{equation}
    S_{J_q}^{^3\text{P}_M} = R^2 \int\dd\bm{r} \abs{\tilde{J}^M_q(\bm{r})}^2,
\end{equation}
where $\tilde{J}^M_t$ corresponds to the $^3\text{P}_M$ components of the pairing spin-current.
The scale factor $R$ is fixed to unity in the present study.

\subsection{Extension for finite-magnetic field systems}
The HF and HFB frameworks can be extended to systems under strong magnetic field by taking into account two distinct effects:
(i) the modification of the nuclear single-particle Hamiltonian, and (ii) the Landau quantization of background electrons.
The detailed formalism has been presented in our previous work~\cite{yoshimura2025}, and here we briefly summarize the essential ingredients relevant to the present study.

The single-particle Hamiltonian in the presence of a magnetic-field is written as
\begin{equation}
    \hat{h}_{q,\sigma\sigma'} = \hat{h}^{(0)}_{q,\sigma\sigma'} +  \hat{h}^{(B)}_{q,\sigma\sigma'},
\end{equation}
where $\hat{h}^{(0)}_{q,\sigma\sigma'}$ denotes the Hamiltonian in the absence of magnetic field, while $\hat{h}^{(B)}_{q,\sigma\sigma'}$ represents the magnetic contribution.
The latter consists of orbital and spin couplings,
\begin{equation}
    \hat{h}^{(B)}_{q,\sigma\sigma'} = -\qty(\bm{l}\delta_{qp} + g_q\frac{\bm{\sigma}}{2}) \bm{\cdot}\tilde{\bm{B}}_q.\label{Eq:hB_normal}
\end{equation}
where $\bm{l}$ is the orbital angular momentum operator and $\bm{\sigma}$ denotes the Pauli matrices.
The orbital term acts only on protons, while the second term describes the Zeeman coupling of the nucleon spin to the magnetic field.
In the present work, we neglect the orbital coupling of protons, since our primary interest lies in magnetic effects on neutron superfluidity and pairing correlations, and the orbital contribution is expected to play a minor role.
The nucleon $g$-factors are given as
\begin{equation}
    g_n = -3.826,\quad g_p = +5.585.
\end{equation}
The dimensionless magnetic field entering the Hamiltonian is defined as
\begin{equation}
    \tilde{\bm{B}}_q
    =
    \frac{e\hbar}{2m_q c}\,\bm{B}.
\end{equation}
Since we are concerned with extremely strong magnetic fields, it is customary to introduce a scaled magnetic field as
\begin{equation}
    \bm{B}_\star \equiv \frac{\bm{B}}{B_c},
\end{equation}
where $B_c$ is the critical magnetic field at which the electron cyclotron energy becomes comparable to the electron rest mass,
\begin{equation}
    B_c = \frac{m_e^2c^3}{eh} \simeq 4.41\times 10^{13}\,\text{G}.
\end{equation}
In principle, the magnetic field is a three-dimensional vector.
However, owing to the symmetry of the rod phase, the $x$ and $y$ directions are equivalent. Accordingly, we consider magnetic fields oriented either along the $x$ direction (perpendicular to the rods) or along the $z$ direction (parallel to the rods), in the present study.

The Landau energy levels of relaticistic electrons in a magnetic-field are given by
% The energy of Landau levels of relativistic electrons is given by
\begin{equation}
    e_\nu = \sqrt{c^2p^2 + m_\text{e}^2c^4(1+2\nu \abs{B_\star})},
\end{equation}
where $\nu$ is a non-negative integer labeling the Landau level and $p_z$ is the electron momentum parallel to the magnetic field.
For a fixed electron chemical potential $\mu_e$, the allowed Landau levels satisfy
\begin{equation}
    \nu \leq \frac{1}{2\abs{B_\star}}\qty(\frac{\mu_\text{e}^2}{m_\text{e}^2c^4} - 1).\label{eq:2-C:nucond}
\end{equation}
Within this range, the electron number density and energy density are expressed as
\begin{eqnarray}
    n_\text{e} &=& \frac{2\abs{B_\star}}{(2\pi)^2\lambda_\text{e}^3} \sum_{\nu} g_\nu x_\text{e}(\nu),\\
    \mathcal{E}_\text{e} &=& \frac{\abs{B_\star} m_\text{e}c^2}{(2\pi)^2\lambda_\text{e}^3} \sum_{\nu} g_\nu(1+2\nu \abs{B_\star}) \psi_+ \qty[\frac{x_\text{e}(\nu)}{\sqrt{1+2\nu \abs{B_\star}}}] - n_\text{e}m_\text{e}c^2,
\end{eqnarray}
with
\begin{equation}
    \begin{aligned}
        \psi_\pm(x) &= x\sqrt{1+x^2} \pm \ln(x+\sqrt{1+x^2}),&\gamma_\text{e} &= \frac{\mu_\text{e}}{m_\text{e}c^2},\\
        x_\text{e}(\nu) &= \sqrt{\gamma_\text{e}^2 - 1 - 2\nu \abs{B_\star}},&\lambda_\text{e} &= \frac{\hbar}{m_\text{e}c}.
    \end{aligned}
\end{equation}
The electron chemical potential $\mu_e$ is determined self-consistently by imposing both the $\beta$-equilibrium condition,
$\mu_n = \mu_p + \mu_e$, and the charge neutrality condition,
\begin{equation}
    \frac{1}{L_x L_y}
    \int \dd x\,\dd y\,n_p(\bm{r})
    =
    n_e .
\end{equation}

The degree of spin-polarization for nucleons of species $q$ is quantified by
\begin{equation}
    P_{q} = \frac{N_{q,\up} - N_{q,\down}}{N_{q,\up} + N_{q,\down}},
\end{equation}
where the numerator is equivalent to
\begin{equation}
    N_{q,\up} - N_{q,\down} = \int\dd\bm{r}\, \qty[n_{q,\up}(\bm{r}) - n_{q,\down}(\bm{r})] = \int\dd\bm{r}\,s_{z,q}(\bm{r}).\label{eq:spinpolarization}
\end{equation}
The spin coupling to the magnetic field introduces an additional contribution to the energy density functional,
\begin{equation}
    \mathcal{E}_\text{mag}(\bm{r}) = \sum_{q=n,p}-\mu_Bg_q \bm{s}_q\cdot \bm{B}_\star,
\end{equation}
which explicitly shows that the magnetic-field couples to the corresponding components of the spin density.

%% file: 3-setting.tex
\section{COMPUTATIONAL SETTINGS}
\subsection{General settings}
In this study, we develop a dedicated numerical code to perform superfluid band calculations for two-dimensional crystalline structures.
Assuming translational invariance along the $z$ direction, the spatial domain in the $x$ and $y$ directions is discretized on a uniform grid with $N_x = N_y = 24$ points and a grid spacing of $\Delta x = \Delta y = 1~\mathrm{fm}$.
Accordingly, the size of the unit cell is fixed to $L_x = L_y = 24~\mathrm{fm}$ throughout this work.
In the momentum space, the plane-wave wavenumbers are discretized into one hundred points with a spacing of $\Delta k = 0.015~\mathrm{fm}^{-1}$ up to a cutoff $k_{\mathrm{max}} = 1.5~\mathrm{fm}^{-1}$.
For the Bloch wave vectors, the two-dimensional Brillouin zone is discretized using a uniform mesh of $N_{k_x} = N_{k_y} = 32$ points.
A square lattice configuration is assumed for the crystalline structure.
Single-particle wave functions are obtained by diagonalizing the HF or HFB Hamiltonian, and the calculations are iterated self-consistently until convergence of the total energy is achieved.
Spatial derivatives and the Poisson equation are evaluated using the fast Fourier transform (FFT) algorithm.
For the update of densities at each iteration step, the modified Broyden method is adopted.
We use the SLy4~\cite{chabanat1997, chabanat1998} parameter set for the Skyrme EDF.
The baryon density is fixed at $n_B = 0.04~\mathrm{fm}^{-3}$, and the $\beta$-equilibrium condition described in the previous section is imposed self-consistently at each iteration step.
Regarding the coupling constants of the pairing spin-current terms, there is presently no consensus on their appropriate values.
In the present study, for simplicity and to isolate qualitative effects, we adopt identical coupling constants for all tensor components, i.e., $\tilde{C}_t^{J0} = \tilde{C}_t^{J1} = \tilde{C}_t^{J2} = 60\,\mev$.

\subsection{Structure of study}
This study addresses several closely related issues: namely,
(i) the entrainment effect under strong magnetic-fields,
(ii) the combined influence of spin-orbit coupling and magnetic fields on spin polarization and pairing phases, and
(iii) the role of spin-triplet pairing interactions in the presence of magnetic-fields.
Accordingly, different forms of the Hamiltonian are employed in different parts of the analysis.
In the first place, the entrainment effect is investigated within the HF framework, using the Hamiltonian given in Eq.~\eqref{eq:SkyrmeHF}, where pairing correlations are neglected.
The neutron effective mass is defined through Eqs.~\eqref{eq:effectivemass}--\eqref{eq:mobilitycoefficient}, and its dependence on the magnetic-field strength is systematically examined.
In the second place, superfluid effects are included to study the response of spin-polarization and spin-singlet pairing to external magnetic-fields. 
In this part, two calculation settings are considered: one including the spin-orbit interaction and the other neglecting it, in order to clarify its role.
In the third place, the emergence of spin-triplet superfluidity is investigated by performing calculations both with and without the spin-triplet pairing energy density functional.
For each configuration, calculations are carried out over a range of magnetic-field strengths to explore how the condensation of spin-triplet components depends on external conditions.

These calculations require substantial computational resources.
For the HF calculations, CPU parallelization is implemented over the Bloch wave vectors and plane-wave momenta $\bm{k}$.
In the HFB calculations, the computational cost is dominated by the diagonalization of large $4N \times 4N$ Hamiltonian matrices.
To accelerate this procedure, we employ GPU parallelization using the
\texttt{CUDA} framework for Hermitian matrix operations.
To reduce the overall computational cost, a part of the calculations is first performed without explicitly including the Bloch wave vectors $k_x$ and $k_y$.
The impact of fully incorporating band-structure effects is examined and discussed in the later sections of this paper.

%% file: 4-result.tex
\section{RESULTS and DISCUSSION}
In this section, we present and discuss the numerical results, organized according to their physical implications.
First, we investigate the impact of the magnetic field on the neutron effective mass within the HF framework, excluding pairing correlations. 
Secondly, we include superfluidity to the response of the spin-singlet pairing and spin-polarization to magnetic-fields. 
Finally, we analyze how spin-triplet condensations are modified by the presense of the spin-triplet pairing interactions and magnetic fields.

\subsection{Entrainment effects}
\begin{table}[b]
    \centering
    \caption{Calculated free neutron density, condutcion number density and neutron effective mass under several magnetic-field strengths $\abs{\bm{B}_\star}= 0$, $100$, $1000$. For the convenience of comparison, values are scaled by the average neutron density or neutron bare mass.}
    \begin{tabular*}{0.8\columnwidth}{@{\extracolsep{\fill}}lccccc}
        \hline\hline
         & $\bm{B}=\bm{0}$ & $B_x=100$ & $B_x=1000$ & $B_z=100$ & $B_z=1000$ \\
        \hline
        $n^{f}_{n} / n_{n}$ 
        & 0.881 & 0.881 & 0.881 & 0.881 & 0.880 \\
        $n^{c}_{n} / n_{n}$ 
        & 0.586 & 0.577 & 0.380 & 0.578 & 0.389 \\
        $m^{\star}_{n} / m_{n}$ 
        & 1.504 & 1.526 & 2.317 & 1.523 & 2.262 \\
    \hline\hline
    \label{tab:Entrainment}
    \end{tabular*}
\end{table}

\begin{figure}[tp]
    \centering
    \includegraphics[width=0.5\columnwidth]{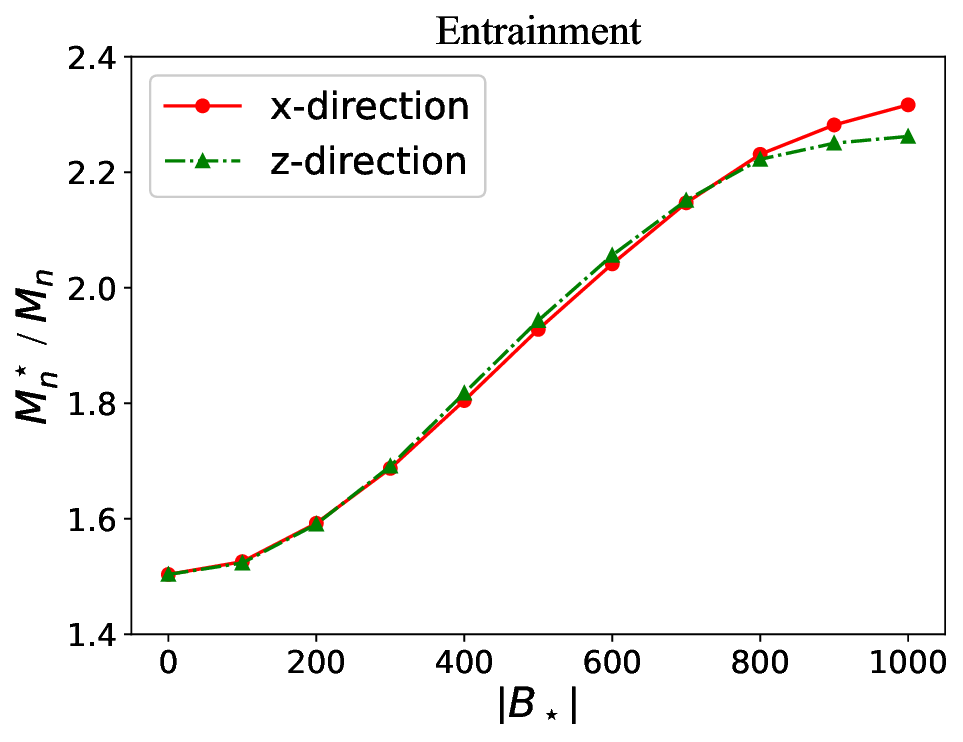}
    \caption{Calculated neutron effective masses as a function of magnetic-field strength. The red solid (green dash-dotted) line corresponds to the results when the magnetic direction is along $x$-axis ($z$-axis) perpendicular (parallel) to the rods.}
    \label{fig:MagEnt}
\end{figure}

\subsubsection{Effective mass}

Table~\ref{tab:Entrainment} summarizes the calculated conduction number density and effective mass for several magnetic field strengths. 
Calculations are performed for two magnetic-field orientations: along the $x$-axis, perpendicular to the nuclear rods, and along the $z$-axis, parallel to the rods.
The dimensionless magnetic-field strengths $B_\star=100$ and $1000$ correspond to approximately $4.4\times10^{15}$~G and $4.4\times10^{16}$~G, respectively, both of which are within the expected range for magnetars.
This table indicates that, in the absence of magnetic field, the neutron effective mass is approximately $m^\star_n/m_n\simeq 1.5$, indicating a clear entrainment regime.
This result is consistent with the band-theoretical calculations by Carter \textit{et al.}~\cite{carter2005}, whereas it contrasts with previous one-dimensional studies, including Ref.~\cite{kashiwaba2019, sekizawa2022} and our earlier work~\cite{yoshimura2024a}, where anti-entrainment was observed.
However, it remains unclear how these results might change when superfluidity is explicitly included or when the effective mass is derived via time-evolution calculations; thus, this point remains open for further discussion.
Regarding the magnetic-field dependence, the entrainment effect remains almost unchanged for moderate field strengths around $B_\star=100$.
In contrast, at $B_\star=1000$, the effective mass exceeds twice the bare neutron mass, corresponding to an enhancement by a factor of approximately 1.5 compared to the zero magnetic-field case.
Figure~\ref{fig:MagEnt} shows the effective mass as a function of magnetic-field strength, demonstrating a smooth increase from $m_n^\star/m_n \simeq 1.4$ to values exceeding 2.0 over the explored range.
Notably, this behavior is essentially independent of the magnetic-field orientation, with similar trends observed for both $x$- and $z$-directions.
As mentioned earlier, $B_\star=100$--$1000$ corresponds to approximately $10^{15}$--$10^{16}$ G in cgs units, a magnitude expected to exist within magnetars. 
These findings suggest that conventional frameworks for astrophysical dynamics, may not be applicable to magnetars in their current form. 
Consequently, this implies the necessity for more detailed future investigations into the internal structure and dynamics of these systems in the presence of strong magnetic fields.

\begin{figure}[tp]
    \centering
    \includegraphics[width=1.0\columnwidth]{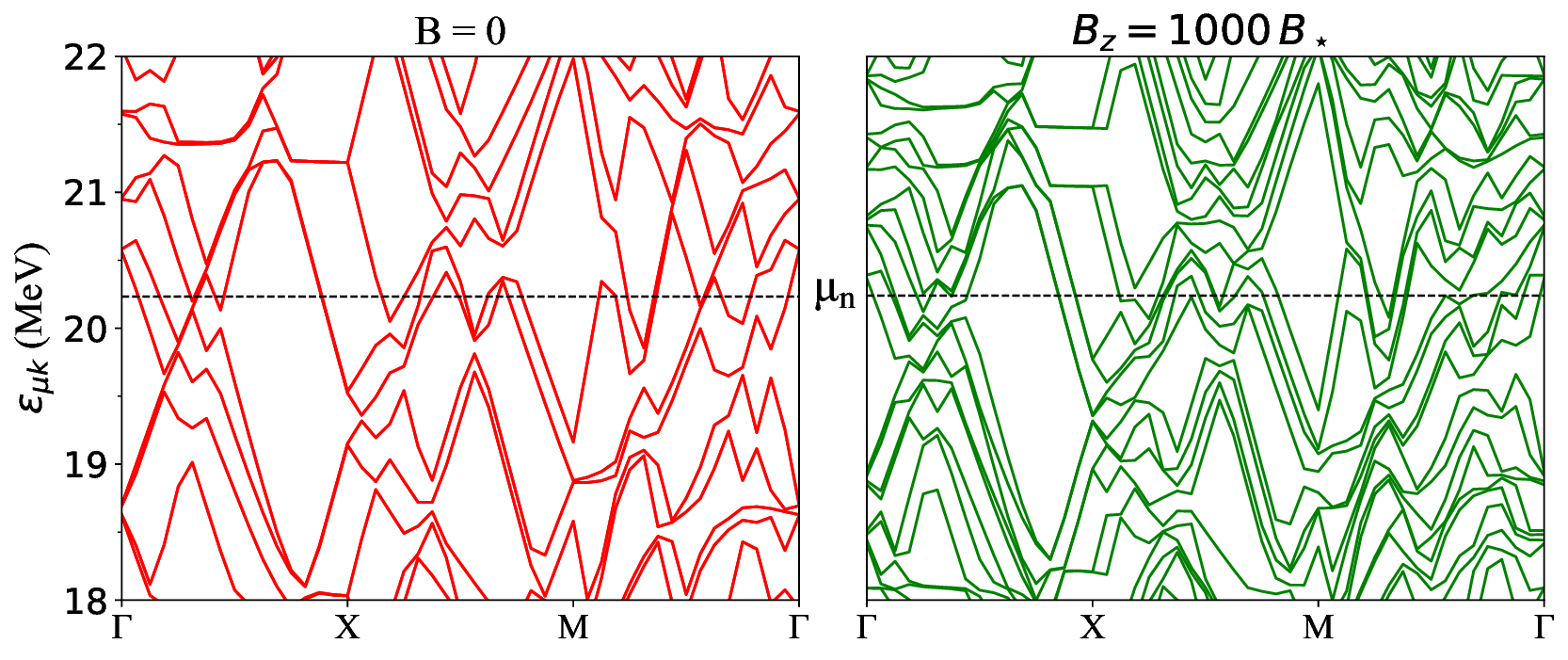}
    \caption{The energy band diagram along a representative path in the Brillouin zone of the square lattice. The left figure shows the results of zero magnetic-field case, whereas the right is for $B_z = 1000B_\star$ case. The black dashed line in two figures indicate the chemical potentials of neutrons.}
    \label{fig:BandFigure}
\end{figure}

\subsubsection{Band diagram}

Figure~\ref{fig:BandFigure} presents the calculated band structures of neutrons along a representative path in the Brillouin zone for the square lattice, both in the absence of a magnetic field (left panel) and in the presence of a magnetic field of $B_z = 1000 B_\star$ (right panel). 
In this figure, a clear enhancement of entrainment can be observed under strong magnetic fields, which can be traced back to qualitative changes in the band structure.
In the zero magnetic-field case, the energy bands are doubly degenerate owing to Kramers degeneracy.
When a magnetic field is applied, this degeneracy is lifted by the Zeeman effect, leading to proliferation of distinct energy bands near the Fermi surface.
The resulting dense spectrum promotes band flattening through level repulsion, which in turn enhances the neutron effective mass.
A simple estimate indicates that the Zeeman splitting at $B_\star=1000$ is of the order of $0.5$~MeV, exceeding the typical band spacing near the Fermi level.
This leads to a highly intricate band structure in that energy region and provides a microscopic explanation for the substantial increase in the effective mass.
Thus, by examining the detailed band structure, we explicitly demonstrate how strong magnetic fields modify macroscopic transport properties such as the neutron effective mass.
% The reason is that the number of visible bands is significantly larger in the presence of the magnetic field compared to the zero-field case. 
% This difference arises because, in the absence of a magnetic field, the bands are degenerate due to Kramers degeneracy, whereas under a finite magnetic field, this degeneracy is lifted by the Zeeman effect, causing the bands to split. 
% The presence of a dense spectrum of energy bands tends to induce band flattening via level repulsion; consequently, it is surmised that this mechanism leads to an increase in the effective mass.
% Indeed, a simple estimation suggests that the Zeeman splitting magnitude at $B_\star=1000$ is approximately $0.5$ MeV. 
% This value exceeds the typical inter-band spacing near the Fermi surface, resulting in a highly intricate band structure in that region. 
% Therefore, by examining the microscopic energy structure, we can explicitly visualize how the magnetic field modifies the macroscopic effective mass.

\subsection{Spin-orbit couplings}
In this section, we present and discuss the results obtained from HFB calculations that explicitly incorporate pairing correlations. For the sake of computational simplicity, band-structure effects are neglected in this subsection by setting $N_{k_x}=N_{k_y}=1$. The influence of band-structure effects will be examined separately in a subsequent section.

\begin{figure}[tp]
    \centering
    \includegraphics[width=1.0\columnwidth]{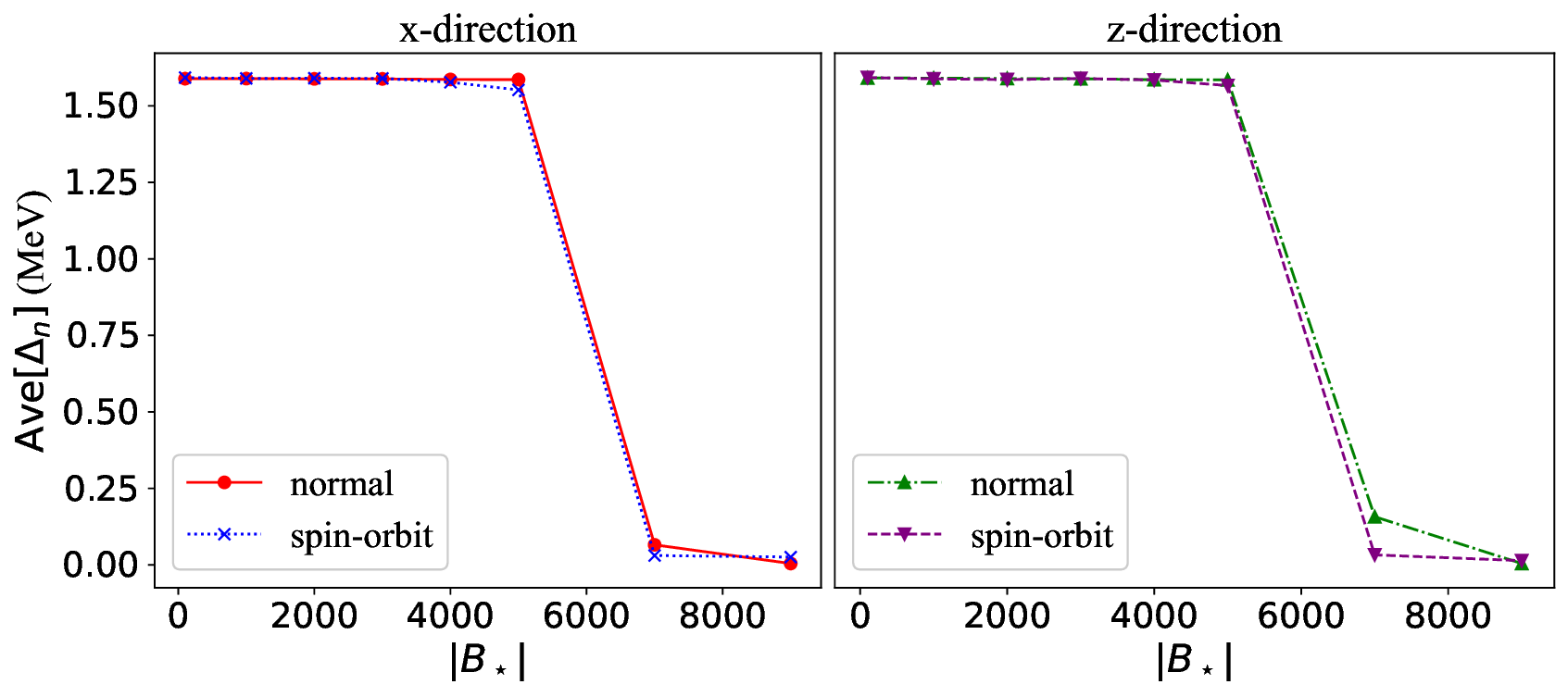}
    \caption{The averaged pairing gaps of spin-singlet neutron superfluidity for a range of magnetic-field strengths. The left figure shows the result for the case where the magnetic-field is imposed along $x$-axis, while the right one is for $z$-axis case. In both figures, the results both when the spin-orbit EDF is present and absent is demonstrated.}
    \label{fig:Delta_singlet}
\end{figure}

Figure \ref{fig:Delta_singlet} plots the averaged pairing gap of the spin-singlet neutron superfluid as a function of magnetic field strength. 
The results are displayed for magnetic fields applied along the $x$- and $z$-axes, and for cases with and without the spin-orbit term in the energy density functional.
This figure indicates that the pairing gap remains nearly constant up to a certain threshold of the magnetic field strength, beyond which, it decreases abruptly and vanishes, signaling a transition to the normal phase. 
This qualitative behavior is robust and does not depend on the magnetic-field direction or on the inclusion of the spin-orbit interaction.

\begin{figure}[tp]
    \centering
    \includegraphics[width=1.0\columnwidth]{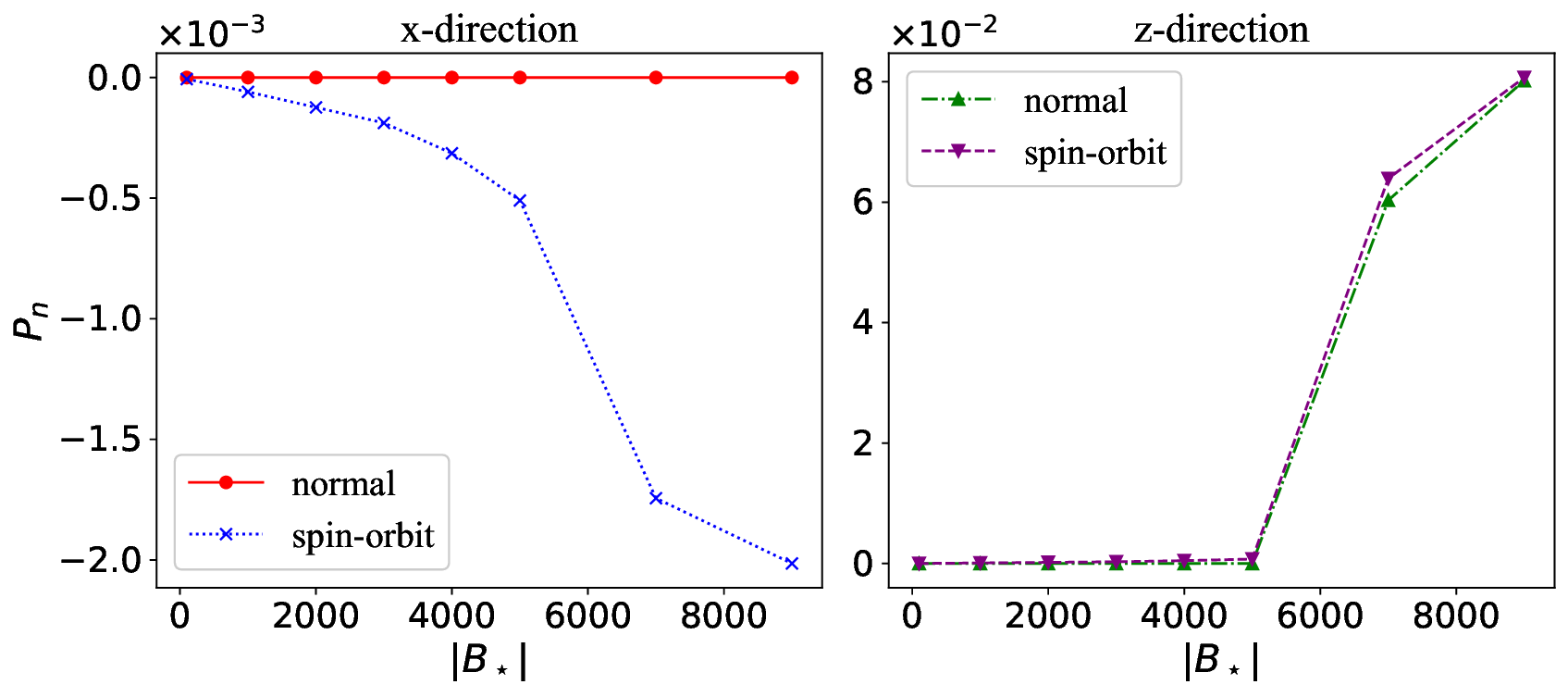}
    \caption{The same with Fig.~\ref{fig:Delta_singlet}, but for the spin-polarization profiles.}
    \label{fig:Polarization}
\end{figure}

Figure \ref{fig:Polarization} plots the magnitude of the neutron spin-polarization as a function of the magnetic field strength. 
Results for magnetic fields applied along the $x$- and $z$-axes are shown in separate panels, with calculations performed both with and without the spin-orbit term.
% In each panel, we display results obtained both with and without the inclusion of the spin-orbit term in the energy density functional (EDF). 
Note that the vertical scales differ between the two panels, being of order $10^{-3}$ for the $x$-axis case and $10^{-2}$ for the $z$-axis case.
Two key features can be drawn from these results. 
In the fitst place, in the absence of the spin-orbit interaction, the spin-polarization is entirely determined by the magnetic-field direction. 
No polarization appears for magnetic fields applied along the $x$-axis, whereas for fields along the $z$-axis the polarization remains zero up to a threshold field strength of approximately $5000\,B_\star$, above which it emerges abruptly.
This behavior is naturally understood from the definition of the spin-polarization in terms of the $z$-component of the spin density [Eq.~\eqref{eq:spinpolarization}] and from the fact that the Zeeman coupling acts directly on this component.
This result is also consistent with our previous findings~\cite{yoshimura2025}.
% Specifically, no polarization occurs when the magnetic field is applied along the $x$-axis. 
% In the case of the $z$-axis, the polarization remains zero up to a certain threshold—specifically, above $5000 B_\star$—at which point it emerges abruptly. 
% This behavior is readily understood from the fact that the spin polarization is defined by the $z$-component of the spin density, as shown in Eq.~\eqref{eq:spinpolarization}, and the Zeeman effect couples directly to the respective components of the spin density.
% This conclusion is also consistent with the findings of our previous study~\cite{yoshimura2025}.
In the second place, the inclusion of the spin-orbit interaction qualitatively modifies the polarization behaviour.
Namely, a finite, albeit small, spin-polarization emerges even when the magnetic field is applied along the $x$-axis, with an onset at much lower field strengths of the order of $1000\,B_\star$.
% The nature of this polarization differs from that observed under a $z$-axis magnetic field; specifically, the onset of polarization occurs at much lower field strengths, around $1000 B_\star$. 
% This behavior was not observed in previous one-dimensional calculations, as the spin-orbit force involves the curl (rotation) of a vector and thus yields no effective contribution in one dimension. 
% Consequently, this result was obtained for the first time by applying our theoretical framework to systems with two or more spatial dimensions.
This behavior was not observed in previous one-dimensional calculations, where the spin-orbit interaction yields no contribution in the slab phase.
The present results therefore demonstrate, for the first time, that the interplay between dimensionality and spin-orbit coupling can induce spin-polarization in superfluid neutron matter under magnetic fields.

\subsection{Spin-triplet pairing}
\begin{figure}[tp]
    \centering
    \includegraphics[width=1.0\columnwidth]{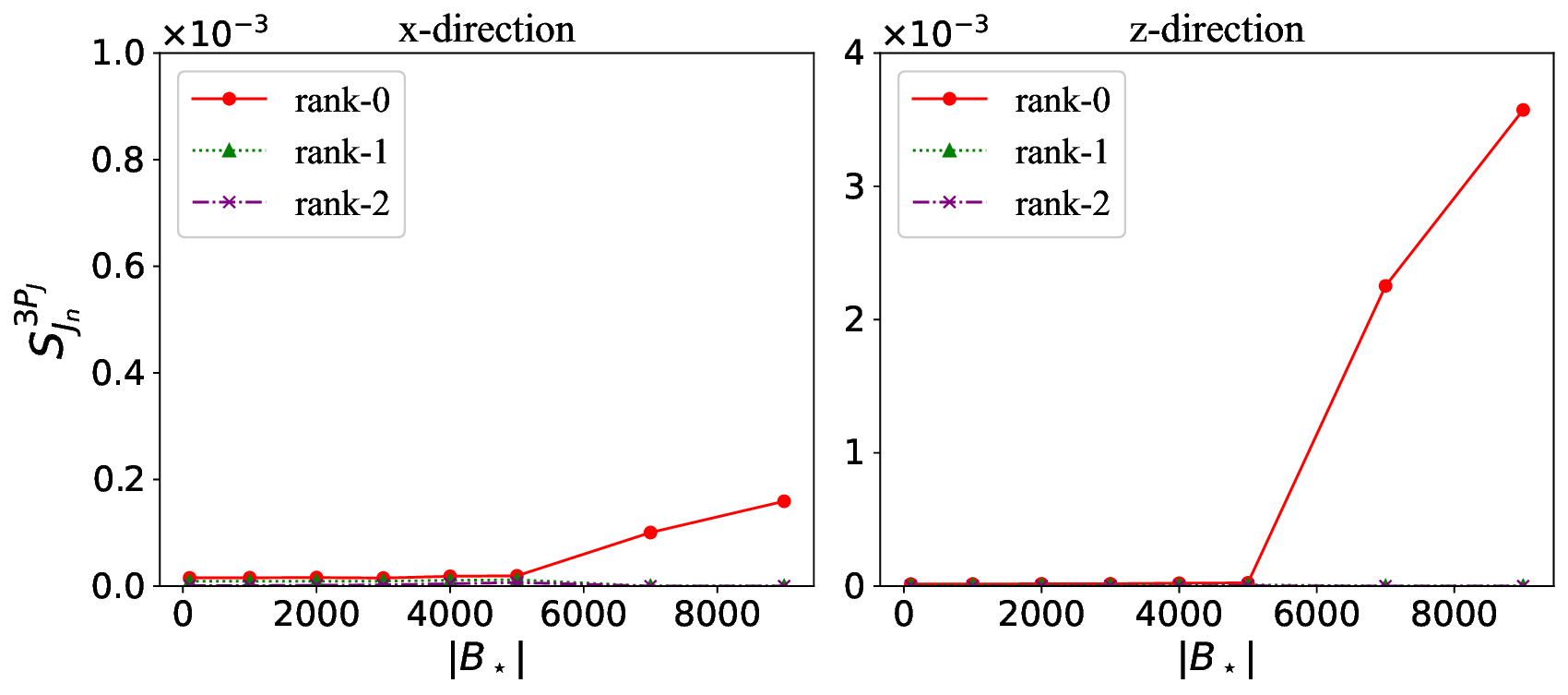}
    \caption{The pairing condensation of each component of spin-triplet neutron superfluidity. In the left figure, the result when the magnetic-field is imposed along $x$-axis is shown, while the right figure is for the $z$-axis case. In both figures, the red solid line, green dotted line, violet dash-dotted line correspond to the sequences of rank-0, rank-1 and rank-2 components, respectively.}
    \label{fig:3PJ_singlet}
\end{figure}

\subsubsection{Spin-triplet EDF}

\begin{figure}[tp]
    \centering
    \includegraphics[width=1.0\columnwidth]{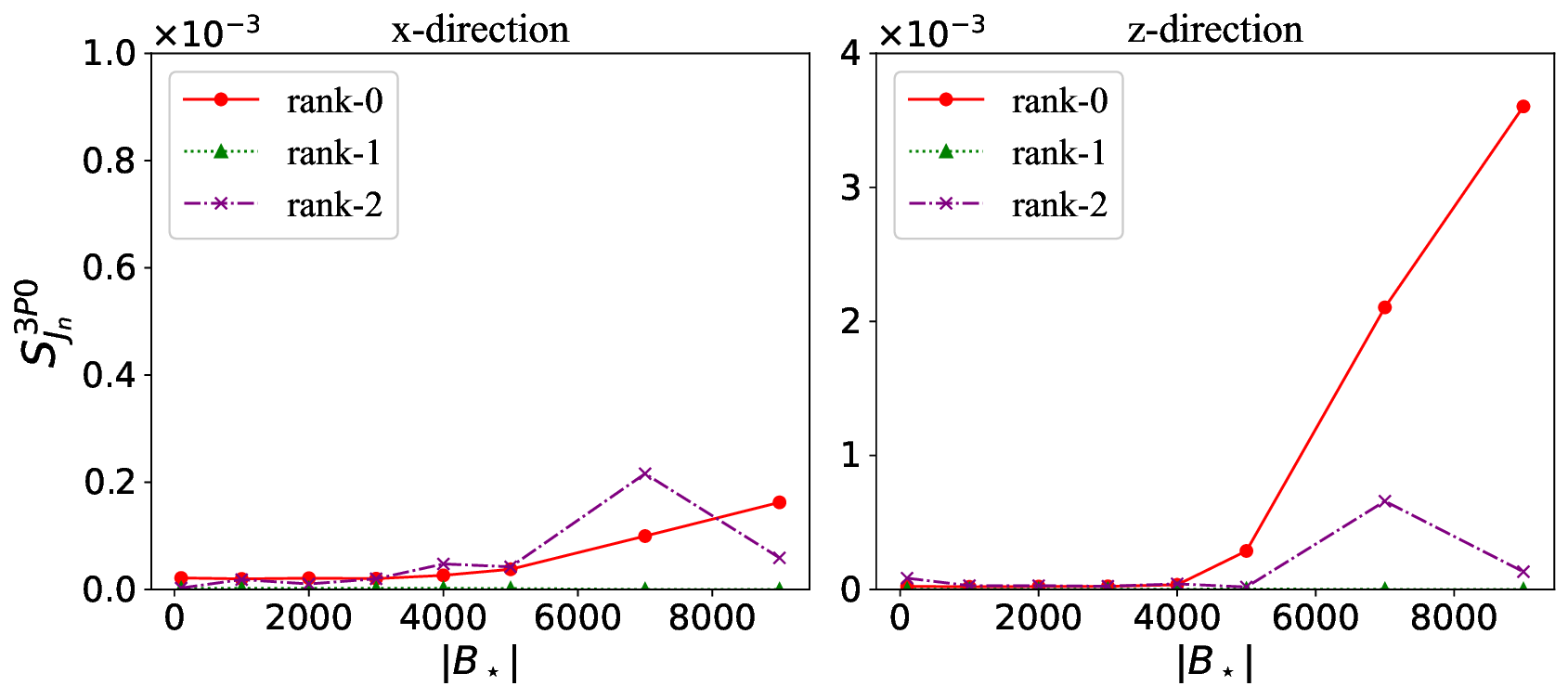}
    \caption{The same with Fig.~\ref{fig:3PJ_singlet}, but for the case where the spin-triplet EDF is incorporated.}
    \label{fig:3PJ_triplet}
\end{figure}

In this section, we discuss the computational results concerning spin-triplet superfluid condensation. 
Figures~\ref{fig:3PJ_singlet} and \ref{fig:3PJ_triplet} show the magnitude of the rank components of the neutron spin-triplet pairing as functions of the magnetic-field strengths.
Figure~\ref{fig:3PJ_singlet} corresponds to calculations without the spin-triplet pairing EDF, while Fig.~\ref{fig:3PJ_triplet} presents the results obtained when this interaction is included.
By comparing these two figures, three conclusions can be drawn.
In the first place, in both calculations, the rank-0 component emerges abruptly once the magnetic-field strength exceeds a critical value, reaching magnitudes of the order of $10^{-3}$.
This threshold takes place around $B_\star = 5000$, which coinsides with the critical magnetic field for the appearance of spin polarization.
This correspondence indicates that the emergence of the rank-0 component is closely linked to the spin-polarized phase, consistent with previous studies highlighting the relationship between polarized neutron matter and the $^3\text{P}_0$ pairing order parameter~\cite{tajima2023}.
In the second place, the rank-2 component never appears in the absense of the spin-triplet pairing EDF, whereas it acquires finite values whenever the corresponding interaction channel is included, largely independent of the magnetic-field strength.
This result demonstrates that the $^3\text{P}_2$ pairing component is driven primarily by the presense of the spin-triplet interaction itself, in contrast to the $^3\text{P}_0$ component, which is induced by spin polarization.
In the third place, the rank-1 component remains negligibly small in all cases, irrespective of the magnetic-field direction, strength, or the inclusion of the spin-triplet pairing EDF.
Since the rank-1 component corresponds to the vector part of the pairing spin current, which is forbidden by spatial inversion symmetry, its suppression can be attributed to the inversion symmetry of the two-dimensional rod phase around the $z$-axis.

\subsubsection{Band structure effects}
\begin{table}[b]
    \centering
    \caption{The calculation results of rank-0 and rank-2 components of spin-triplet neutron superfluidity, for both cases where the band structure effects are present and absent. In all cases, results for $B_\star = 1000$ and $5000$ are demonstrated.}
    \begin{tabular*}{0.9\columnwidth}{@{\extracolsep{\fill}}c c c | c c c}
        \hline
        \multicolumn{3}{c|}{\textbf{Bx}} & \multicolumn{3}{c}{\textbf{Bz}} \\
        \hline
        \multicolumn{6}{c}{} \\[-1.2ex]
        \multicolumn{3}{c|}{\textbf{3P0}} & \multicolumn{3}{c}{\textbf{3P0}} \\
        \hline
        $\abs{B}/B_\star$  & No Band & Band & $\abs{B}/B_\star$  & No Band & Band\\
        \hline
        1000 & $9.13\times10^{-5}$ & $8.34\times10^{-5}$ & 
        1000 & $8.51\times10^{-5}$ & $8.53\times10^{-5}$ \\
        5000 & $1.64\times10^{-4}$ & $2.06\times10^{-4}$ & 
        5000 & $2.67\times10^{-4}$ & $2.34\times10^{-4}$ \\
        \hline
        \multicolumn{6}{c}{} \\[-1.2ex]
        \multicolumn{3}{c|}{\textbf{3P2}} & \multicolumn{3}{c}{\textbf{3P2}} \\
        \hline
        $\abs{B}/B_\star$  & No Band & Band & $\abs{B}/B_\star$  & No Band & Band \\
        \hline
        1000 & $1.85\times10^{-5}$ & $4.38\times10^{-5}$ & 
        1000 & $6.23\times10^{-6}$ & $8.38\times10^{-6}$ \\
        5000 & $4.23\times10^{-5}$ & $2.47\times10^{-4}$ & 
        5000 & $4.17\times10^{-5}$ & $4.31\times10^{-5}$ \\
        \hline
        \label{tab:3PJ_band}
    \end{tabular*}
\end{table}

% The results presented thus far have entirely neglected the effects of band wavenumbers. In the following, we examine the extent to which band structure effects influence spin-triplet pairing.
Thus far, the results have been obtained without explicitly including Bloch wavenumbers.
We now examine how band-structure effects influence spin-triplet pairing.

Table~\ref{tab:3PJ_band} summarizes the calculated rank-0 and rank-2 components with and without band-structure effects for representative magnetic-field strengths.
From this table, it is found that the qualitative dependence of each component on the magnetic-field remains essentially unchanged when band structure effects are included, although their magnitudes are quantitatively modified.
% On the contrary, only the $^3P_2$ components when the magnetic-fields are along $x$-axis are different by the magnitude order.
% Nevertheless, this does not change our main statement, and consequently, we conclude that the emergence of spin-triplet superfluidity is governed primarily by the spin-polarization phases or the interaction channnels, which is robust regardless of the crystalline structure effects.
An exception is found for the $^3\text{P}_2$ component under magnetic fields applied along the $x$-axis, where band-structure effects lead to an enhancement by approximately one order of magnitude.
Nevertheless, this enhancement does not alter the overall conclusion: the emergence of spin-triplet superfluidity is governed primarily by spin-polarization and by the presence of specific interaction channels, and remains qualitatively robust against crystalline band-structure effects.

% Table \ref{tab:3PJ_band} presents the values of the rank-0 and rank-2 components calculated with and without band wavenumbers for several magnetic field strengths. It is evident from the table that, although the inclusion of band structure effects introduces slight quantitative differences in magnitude, the qualitative behavior with respect to the magnetic field remains largely unchanged compared to the case where these effects are neglected.
%
% Consequently, we conclude that the emergence of spin-triplet superfluidity is governed primarily by the magnitude and direction of the magnetic field and the nucleon-nucleon interaction channels, while the effects of the crystal structure play only a minor role.

%% file: 5-summary.tex
\section{SUMMARY AND PROSPECTS}
In this work, we have extended the framework of the superfluid band theory, developed in our previous studies, to the two-dimensional rod phase of the neutron star inner crust. Moreover, we have implemented spin-triplet pairing into the self-consistent superfluid band theory.
For describing the spin-triplet neutron superfluidity, the $S=1$ component of the anomalous density is incorporated in the superfluid DFT, on the same footing with the effects of finite magnetic-fields.
Within this unified framework, we have systematically investigated spin-polarization and pairing phases by selectively switching on and off the spin-orbit interaction and spin-triplet pairing EDF.

Our main findings can be summarized as follows.
In the first place, on the magnetic-fields on the order of $10^{16}$~G, the effective masses of neutrons increase by a factor of around $1.5$ compared to no magnetized cases.
This enhancement originates from a sizable modification of the band structure when the Zeeman splitting becomes comparable to the inter-level spacing.
The result indicates that the conventional application of band theory to crustal matter may become questionable in magnetars, and calls for a more careful microscopic treatment of neutron dynamics under such extreme conditions.
In the second place, we have examined the response of spin polarization and pairing gaps to magnetic fields, with particular attention to the role of the spin-orbit interaction.
% The obtained results demonstrate that while the behaviour of the pairing gap depends solely on the magnitude of magnetic-fields, the contribution of the spin-orbit term allows for the emergence of slightly polarized phases, even when the magnetic-field direction is neither so large nor aligned with the spin-quantization axis.
While the pairing gap has been found to depend primarily on the magnitude of the magnetic field, the spin-orbit interaction enables the emergence of weakly spin-polarized phases even for moderate field strengths and non-aligned field directions.
% In conventional studies of neutron star crust matter, the spin-orbit effect has often been considered minor compared to other terms and has thus not been deeply explored. 
% Besides, previous superfluid band calculations were limited to one-dimensional systems, where the spin-orbit term yields no finite contribution. 
Such effects have not been captured in previous superfluid band calculations, which were restricted to one-dimensional geometries where spin-orbit contributions vanish.
Our spin-dependent HFB calculations in the two-dimensional rod phase thus reveal, for the first time, the nontrivial interplay between dimensionality, spin-orbit coupling, and magnetic fields in crustal matter.
% The present results have been elucidated for the first time by performing spin-dependent HFB calculations for the two-dimensional rod phase.
% In the third place, we have systematically investigtated the emergence of spin-triplet superfluid components, which is a topic that has recently attracted attention in the context of glitch dynamics.
% From the calculation results it has been revealed that the rank-0 component emerges solely due to the spin-polarization brought by the magnitude of the magnetic-field, whereas the rank-2 component appears by the effects of the specific interaction channel.
% Furthermore, by comparing results with and without band structure effects, we have also demonstrated that these superfluid properties remain robust under the influence of crystalline structures, even though their condensations may quantitatively differ.
% The self-consistent nuclear calculations for neutron tripled-odd pairing still remain rare; in particular, their application to crustal matter is pioneered in this study.
% This topic requires further development in conjunction with numerical simulations of glitch phenomena.
In the third place, we have systematically investigated the emergence of spin-triplet superfluid components.
The results have demonstrated that the rank-0 component arises solely from magnetic-field-induced spin polarization, whereas the rank-2 component is generated by specific spin-triplet pairing interactions.
By comparing calculations with and without band-structure effects, we further showed that these superfluid properties remain qualitatively robust in crystalline environments, although their magnitudes are quantitatively modified.
Self-consistent nuclear calculations of neutron spin-triplet pairing remain scarce, and their application to neutron-star crust matter is pioneered in the present study.
These findings provide a microscopic basis for future investigations of astronomical simulations involving unconventional superfluid phases.

% As further developments, we envision extending the calculations to the time-dependent framework, similar to our previous work, to determine the neutron effective mass with superfluid effects expliticly included.
% Although the superfluid fractions of inner crust phases have been already investigated by several researches, their calculation frameworks are limited in static configurations~\cite{almirante2024, almirante2024a, almirante2025, almirante2025a, chamel2025}.
% One of our interests is in the superfluid dynamics of the nuclear crystals and in this sense above extension is still an important issue.
% Additionally, it is intriguing how the spin-triplet superfluidity and the spin-polarizations are affected within the three-dimensional crystalline structures.
% Although these extensions require overwhelming calculation costs, relying on the recent advancement of computational resources, as well as of the numerical methodologies such as Conjugate-Orthogonal-Conjugate-Gradient (COCG)~\cite{jin2017} and COC-Redisual method~\cite{kashiwaba2020}, they are expected to become possible.
% We hope that through the long journey of further developments in nuclear research, advancements in observational technology, and collaboration with astrophysical simulations, these problems will be fully resolved, leading to a dramatic
% advancement in physics.
As future developments, it is highly desirable to extend the present framework to time-dependent calculations, following our previous work~\cite{yoshimura2024a}, in order to determine neutron effective masses with superfluid effects explicitly included.
Although superfluid fractions in the inner crust have been investigated in several static approaches~\cite{almirante2024, almirante2024a, almirante2025, almirante2025a, chamel2025}, a dynamical description of superfluid motion in nuclear crystals remains largely unexplored.
Another important direction is the extension to fully three-dimensional crystalline structures, where spin-triplet superfluidity and spin polarization may exhibit richer behavior.
While such studies require substantial computational resources, recent advances in numerical methodologies, including the conjugate-orthogonal conjugate-gradient (COCG)~\cite{jin2017} and related techniques~\cite{kashiwaba2020}, together with modern high-performance computing, make these extensions increasingly feasible.
We hope that continued progress in microscopic nuclear theory, observational astronomy, and large-scale astrophysical simulations will ultimately lead to the comprehensive understanding of superfluid phenomena in neutron stars.

%% file: 6-acknowledgement.tex
\section*{ACKNOWLEDGEMENT}
We would like to appreciate Kenichi Yoshida (RCNP, Osaka University) and Nobuo Hinohara (Tsukuba University) for the kind introduction of the formalism of the spin-triplet pairing.
The author K.Y. acknowledges Shunsuke Yasunaga (Institute of Science Tokyo) who checked the formulation of this work from a theoretical point of view.
This work is supported by JSPS Research Fellow, Grant No.~JP24KJ1110, as well as JSPS Grantin-Aid for Scientific Research, Grants No.~JP23K03410, No.~JP23K25864, and No.~JP25H01269
This work mainly used the computational resources of TSUBAME4.0 at Institute of Science Tokyo, through the HPCI System Project, Project ID: hp230180, hp240183, and hp250097.
Additionally this work used computational resources of the Yukawa-21 supercomputer at Yukawa Institute for Theoretical Physics (YITP), Kyoto University. 
This work partly used computational resource of Miyabi-C provided by Multidisciplinary Cooperative Research Program (MCRP) in Center for Computational Science, University of Tsukuba, Project ID: xg25i043.
This work partly used computational resource of Fugaku provided by RIKEN through the HPCI System Research Project (Project ID: hp250288).

%% file: triplet.bib
@article{almirante2024,
  title = {Superfluid Fraction in the Slab Phase of the Inner Crust of Neutron Stars},
  author = {Almirante, Giorgio and Urban, Michael},
  year = 2024,
  journal = {Phys. Rev. C},
  volume = {109},
  pages = {045805},
  publisher = {American Physical Society},
  url = {https://link.aps.org/doi/10.1103/PhysRevC.109.045805}
}

@article{almirante2024a,
  title = {Superfluid Fraction in the Rod Phase of the Inner Crust of Neutron Stars},
  author = {Almirante, Giorgio and Urban, Michael},
  year = 2024,
  journal = {Phys. Rev. C},
  volume = {110},
  pages = {065802},
  publisher = {American Physical Society},
  url = {https://link.aps.org/doi/10.1103/PhysRevC.110.065802}
}

@article{almirante2025,
  title = {Superfluid {{Density}} in {{Linear Response Theory}}: {{Pulsar Glitches}} from the {{Inner Crust}} of {{Neutron Stars}}},
  shorttitle = {Superfluid {{Density}} in {{Linear Response Theory}}},
  author = {Almirante, Giorgio and Urban, Michael},
  year = 2025,
  journal = {Phys. Rev. Lett.},
  volume = {135},
  pages = {132701},
  publisher = {American Physical Society},
  url = {https://link.aps.org/doi/10.1103/mg61-gw93}
}

@misc{almirante2025a,
  title = {Superfluid Fraction in the Crystal Phase of the Inner Crust of Neutron Stars},
  author = {Almirante, Giorgio and Kaskitsi, Theodora and Urban, Michael},
  year = 2025,
  eprint = {2512.18549},
  primaryclass = {nucl-th},
  publisher = {arXiv},
  url = {http://arxiv.org/abs/2512.18549},
  archiveprefix = {arXiv}
}

@article{andersson2012,
  title = {Pulsar {{Glitches}}: {{The Crust}} Is Not {{Enough}}},
  shorttitle = {Pulsar {{Glitches}}},
  author = {Andersson, N. and Glampedakis, K. and Ho, W. C. G. and Espinoza, C. M.},
  year = 2012,
  journal = {Phys. Rev. Lett.},
  volume = {109},
  pages = {241103},
  publisher = {American Physical Society},
  url = {https://link.aps.org/doi/10.1103/PhysRevLett.109.241103}
}

@article{andersson2021,
  title = {A {{Superfluid Perspective}} on {{Neutron Star Dynamics}}},
  author = {Andersson, Nils},
  year = 2021,
  journal = {Universe},
  volume = {7},
  pages = {17},
  publisher = {Multidisciplinary Digital Publishing Institute},
  issn = {2218-1997},
  url = {https://www.mdpi.com/2218-1997/7/1/17},
  copyright = {http://creativecommons.org/licenses/by/3.0/}
}

@article{basilico2015,
  title = {Outer Crust of a Cold Non-Accreting Magnetar},
  author = {Basilico, D. and Arteaga, D. Pe{\~n}a and {Roca-Maza}, X. and Col{\`o}, G.},
  year = 2015,
  journal = {Phys. Rev. C},
  volume = {92},
  pages = {035802},
  publisher = {American Physical Society},
  url = {https://link.aps.org/doi/10.1103/PhysRevC.92.035802}
}

@article{basilico2025,
  title = {Synthesis of Superheavy Elements in the Outer Crust of a Magnetar},
  author = {Basilico, D. and Col{\`o}, G. and {Roca-Maza}, Xavier},
  year = 2025,
  journal = {Phys. Rev. C},
  volume = {112},
  pages = {015801},
  publisher = {American Physical Society},
  url = {https://link.aps.org/doi/10.1103/55y6-6zxx}
}

@article{bonanno2003,
  title = {Mean-Field Dynamo Action in Protoneutron Stars},
  author = {Bonanno, A. and Rezzolla, L. and Urpin, V.},
  year = 2003,
  journal = {A\&A},
  volume = {410},
  pages = {L33-L36},
  publisher = {EDP Sciences},
  issn = {0004-6361, 1432-0746},
  url = {https://www.aanda.org/articles/aa/abs/2003/42/aafg163/aafg163.html},
  copyright = {\copyright{} ESO, 2003}
}

@article{broderick2000a,
  title = {The {{Equation}} of {{State}} of {{Neutron Star Matter}} in {{Strong MagneticFields}}},
  author = {Broderick, A. and Prakash, M. and Lattimer, J. M.},
  year = 2000,
  journal = {ApJ},
  volume = {537},
  pages = {351},
  publisher = {IOP Publishing},
  issn = {0004-637X},
  url = {https://iopscience.iop.org/article/10.1086/309010}
}

@article{bulgac2002,
  title = {Renormalization of the {{Hartree-Fock-Bogoliubov Equations}} in the {{Case}} of a {{Zero Range Pairing Interaction}}},
  author = {Bulgac, Aurel and Yu, Yongle},
  year = 2002,
  journal = {Phys. Rev. Lett.},
  volume = {88},
  pages = {042504},
  publisher = {American Physical Society},
  url = {https://link.aps.org/doi/10.1103/PhysRevLett.88.042504}
}

@article{bulgac2002a,
  title = {Local Density Approximation for Systems with Pairing Correlations},
  author = {Bulgac, Aurel},
  year = 2002,
  journal = {Phys. Rev. C},
  volume = {65},
  pages = {051305},
  publisher = {American Physical Society},
  url = {https://link.aps.org/doi/10.1103/PhysRevC.65.051305}
}

@article{caplan2017,
  title = {Colloquium: {{Astromaterial}} Science and Nuclear Pasta},
  shorttitle = {Colloquium},
  author = {Caplan, M. E. and Horowitz, C. J.},
  year = 2017,
  journal = {Rev. Mod. Phys.},
  volume = {89},
  pages = {041002},
  publisher = {American Physical Society},
  url = {https://link.aps.org/doi/10.1103/RevModPhys.89.041002}
}

@article{carter2005,
  title = {Entrainment Coefficient and Effective Mass for Conduction Neutrons in Neutron Star Crust: Simple Microscopic Models},
  shorttitle = {Entrainment Coefficient and Effective Mass for Conduction Neutrons in Neutron Star Crust},
  author = {Carter, Brandon and Chamel, Nicolas and Haensel, Pawel},
  year = 2005,
  journal = {Nuclear Physics A},
  volume = {748},
  pages = {675--697},
  issn = {0375-9474},
  url = {https://www.sciencedirect.com/science/article/pii/S0375947404011790}
}

@article{chabanat1997,
  title = {A {{Skyrme}} Parametrization from Subnuclear to Neutron Star Densities},
  author = {Chabanat, E. and Bonche, P. and Haensel, P. and Meyer, J. and Schaeffer, R.},
  year = 1997,
  journal = {Nuclear Physics A},
  volume = {627},
  pages = {710--746},
  issn = {0375-9474},
  url = {https://www.sciencedirect.com/science/article/pii/S0375947497005964}
}

@article{chabanat1998,
  title = {A {{Skyrme}} Parametrization from Subnuclear to Neutron Star Densities {{Part II}}. {{Nuclei}} Far from Stabilities},
  author = {Chabanat, E. and Bonche, P. and Haensel, P. and Meyer, J. and Schaeffer, R.},
  year = 1998,
  journal = {Nuclear Physics A},
  volume = {635},
  pages = {231--256},
  issn = {0375-9474},
  url = {https://www.sciencedirect.com/science/article/pii/S0375947498001808}
}

@article{chamel2005,
  title = {Band Structure Effects for Dripped Neutrons in Neutron Star Crust},
  author = {Chamel, Nicolas},
  year = 2005,
  journal = {Nuclear Physics A},
  volume = {747},
  pages = {109--128},
  issn = {0375-9474},
  url = {https://www.sciencedirect.com/science/article/pii/S0375947404009200}
}

@article{chamel2008,
  title = {Physics of {{Neutron Star Crusts}}},
  author = {Chamel, Nicolas and Haensel, Pawel},
  year = 2008,
  journal = {Living Rev. Relativ.},
  volume = {11},
  pages = {10},
  issn = {1433-8351},
  url = {https://doi.org/10.12942/lrr-2008-10}
}

@article{chamel2012,
  title = {Neutron Conduction in the Inner Crust of a Neutron Star in the Framework of the Band Theory of Solids},
  author = {Chamel, N.},
  year = 2012,
  journal = {Phys. Rev. C},
  volume = {85},
  pages = {035801},
  publisher = {American Physical Society},
  url = {https://link.aps.org/doi/10.1103/PhysRevC.85.035801}
}

@article{chamel2012a,
  title = {Properties of the Outer Crust of Strongly Magnetized Neutron Stars from {{Hartree-Fock-Bogoliubov}} Atomic Mass Models},
  author = {Chamel, N. and Pavlov, R. L. and Mihailov, L. M. and Velchev, {\relax Ch}. J. and Stoyanov, {\relax Zh}. K. and Mutafchieva, Y. D. and Ivanovich, M. D. and Pearson, J. M. and Goriely, S.},
  year = 2012,
  journal = {Phys. Rev. C},
  volume = {86},
  pages = {055804},
  publisher = {American Physical Society},
  url = {https://link.aps.org/doi/10.1103/PhysRevC.86.055804}
}

@article{chamel2013,
  title = {Crustal {{Entrainment}} and {{Pulsar Glitches}}},
  author = {Chamel, N.},
  year = 2013,
  journal = {Phys. Rev. Lett.},
  volume = {110},
  pages = {011101},
  publisher = {American Physical Society},
  url = {https://link.aps.org/doi/10.1103/PhysRevLett.110.011101}
}

@article{chamel2017,
  title = {Entrainment in {{Superfluid Neutron-Star Crusts}}: {{Hydrodynamic Description}} and {{Microscopic Origin}}},
  shorttitle = {Entrainment in {{Superfluid Neutron-Star Crusts}}},
  author = {Chamel, N.},
  year = 2017,
  journal = {J Low Temp Phys},
  volume = {189},
  pages = {328--360},
  issn = {1573-7357},
  url = {https://doi.org/10.1007/s10909-017-1815-x}
}

@article{chamel2024,
  title = {Superfluidity and {{Superconductivity}} in {{Neutron Stars}}},
  author = {Chamel, Nicolas},
  year = 2024,
  journal = {Universe},
  volume = {10},
  pages = {104},
  publisher = {Multidisciplinary Digital Publishing Institute},
  issn = {2218-1997},
  url = {https://www.mdpi.com/2218-1997/10/3/104},
  copyright = {http://creativecommons.org/licenses/by/3.0/}
}

@article{chamel2025,
  title = {Superfluid Fraction in the Crystalline Crust of a Neutron Star: {{Role}} of {{BCS}} Pairing},
  shorttitle = {Superfluid Fraction in the Crystalline Crust of a Neutron Star},
  author = {Chamel, N.},
  year = 2025,
  journal = {Phys. Rev. C},
  volume = {111},
  pages = {045803},
  publisher = {American Physical Society},
  url = {https://link.aps.org/doi/10.1103/PhysRevC.111.045803}
}

@article{chau1992,
  title = {Implications of {{3P}} 2 {{Superfluidity}} in the {{Interior}} of {{Neutron Stars}}},
  author = {Chau, H. F. and Cheng, K. S. and Ding, K. Y.},
  year = 1992,
  journal = {Astrophys. J.},
  volume = {399},
  pages = {213},
  publisher = {IOP},
  issn = {0004-637X},
  url = {https://ui.adsabs.harvard.edu/abs/1992ApJ...399..213C}
}

@article{colo2020,
  title = {Nuclear Density Functional Theory},
  author = {Col{\`o}, G.},
  year = 2020,
  journal = {Adv. Phys. X},
  volume = {5},
  pages = {1740061},
  publisher = {Taylor \& Francis},
  issn = {null},
  url = {https://doi.org/10.1080/23746149.2020.1740061}
}

@article{dean2003,
  title = {Pairing in Nuclear Systems: From Neutron Stars to Finite Nuclei},
  shorttitle = {Pairing in Nuclear Systems},
  author = {Dean, D. J. and {Hjorth-Jensen}, M.},
  year = 2003,
  journal = {Rev. Mod. Phys.},
  volume = {75},
  pages = {607--656},
  publisher = {American Physical Society},
  url = {https://link.aps.org/doi/10.1103/RevModPhys.75.607}
}

@article{dobaczewski1984,
  title = {Hartree-{{Fock-Bogolyubov}} Description of Nuclei near the Neutron-Drip Line},
  author = {Dobaczewski, J. and Flocard, H. and Treiner, J.},
  year = 1984,
  journal = {Nuclear Physics A},
  volume = {422},
  pages = {103--139},
  issn = {0375-9474},
  url = {https://www.sciencedirect.com/science/article/pii/0375947484904330}
}

@incollection{duguet2014,
  title = {The {{Nuclear Energy Density Functional Formalism}}},
  booktitle = {The {{Euroschool}} on {{Exotic Beams}}, {{Vol}}. {{IV}}},
  author = {Duguet, T.},
  editor = {Scheidenberger, Christoph and Pf{\"u}tzner, Marek},
  year = 2014,
  pages = {293--350},
  publisher = {Springer},
  address = {Berlin, Heidelberg},
  url = {https://doi.org/10.1007/978-3-642-45141-6_7},
  isbn = {978-3-642-45141-6}
}

@article{fattoyev2017,
  title = {Quantum Nuclear Pasta and Nuclear Symmetry Energy},
  author = {Fattoyev, F. J. and Horowitz, C. J. and Schuetrumpf, B.},
  year = 2017,
  journal = {Phys. Rev. C},
  volume = {95},
  pages = {055804},
  publisher = {American Physical Society},
  url = {https://link.aps.org/doi/10.1103/PhysRevC.95.055804}
}

@article{frieben2012,
  title = {Equilibrium Models of Relativistic Stars with a Toroidal Magnetic Field},
  author = {Frieben, J. and Rezzolla, L.},
  year = 2012,
  journal = {Monthly Notices of the Royal Astronomical Society},
  volume = {427},
  pages = {3406--3426},
  issn = {0035-8711, 1365-2966},
  url = {https://academic.oup.com/mnras/article-lookup/doi/10.1111/j.1365-2966.2012.22027.x}
}

@article{gogelein2008a,
  title = {Nuclear Matter in the Crust of Neutron Stars Derived from Realistic \$\textbackslash mathit\textbraceleft{{NN}}\textbraceright\$ Interactions},
  author = {G{\"o}gelein, P. and van Dalen, E. N. E. and Fuchs, C. and M{\"u}ther, H.},
  year = 2008,
  journal = {Phys. Rev. C},
  volume = {77},
  pages = {025802},
  publisher = {American Physical Society},
  url = {https://link.aps.org/doi/10.1103/PhysRevC.77.025802}
}

@article{granados2025,
  title = {Half-Vortex States in the Rotating Outer Core of Neutron Stars},
  author = {Granados, J. A. Gil and Mateo, A. Mu{\~n}oz and Vi{\~n}as, X.},
  year = 2025,
  journal = {Phys. Rev. C},
  volume = {111},
  pages = {065802},
  publisher = {American Physical Society},
  url = {https://link.aps.org/doi/10.1103/PhysRevC.111.065802}
}

@article{grill2014a,
  title = {Equation of State and Thickness of the Inner Crust of Neutron Stars},
  author = {Grill, Fabrizio and Pais, Helena and Provid{\^e}ncia, Constan{\c c}a and Vida{\~n}a, Isaac and Avancini, Sidney S.},
  year = 2014,
  journal = {Phys. Rev. C},
  volume = {90},
  pages = {045803},
  publisher = {American Physical Society},
  url = {https://link.aps.org/doi/10.1103/PhysRevC.90.045803}
}

@article{hashimoto1984,
  title = {Shape of {{Nuclei}} in the {{Crust}} of {{Neutron Star}}},
  author = {Hashimoto, Masa-aki and Seki, Hironori and Yamada, Masami},
  year = 1984,
  journal = {Prog Theor Phys},
  volume = {71},
  pages = {320--326},
  issn = {0033-068X},
  url = {https://doi.org/10.1143/PTP.71.320}
}

@article{haskell2015,
  title = {Models of Pulsar Glitches},
  author = {Haskell, Brynmor and Melatos, Andrew},
  year = 2015,
  journal = {Int. J. Mod. Phys. D},
  volume = {24},
  pages = {1530008},
  publisher = {World Scientific Publishing Co.},
  issn = {0218-2718},
  url = {https://www.worldscientific.com/doi/abs/10.1142/S0218271815300086}
}

@misc{hattori2025,
  title = {Exploring {{Interplays Between}} \$\textasciicircum 3\textbackslash text\textbraceleft{{P}}\textbraceright\_2\$ {{Neutron Superfluid Vortices}} and \$\textasciicircum 1\textbackslash text\textbraceleft{{S}}\textbraceright\_0\$ {{Proton Fluxtubes}} in the {{Outer Core}} of {{Neutron Stars}}},
  author = {Hattori, Tatsuhiro and Sekizawa, Kazuyuki},
  year = 2025,
  eprint = {2512.22577},
  primaryclass = {nucl-th},
  publisher = {arXiv},
  url = {http://arxiv.org/abs/2512.22577},
  archiveprefix = {arXiv}
}

@article{hellemans2012,
  title = {Tensor Part of the {{Skyrme}} Energy Density Functional. {{III}}. {{Time-odd}} Terms at High Spin},
  author = {Hellemans, V. and Heenen, P.-H. and Bender, M.},
  year = 2012,
  journal = {Phys. Rev. C},
  volume = {85},
  pages = {014326},
  publisher = {American Physical Society},
  url = {https://link.aps.org/doi/10.1103/PhysRevC.85.014326}
}

@article{hinohara2024,
  title = {Triplet-Odd Pairing in Finite Nuclear Systems: {{Even-even}} Singly Closed Nuclei},
  shorttitle = {Triplet-Odd Pairing in Finite Nuclear Systems},
  author = {Hinohara, Nobuo and Oishi, Tomohiro and Yoshida, Kenichi},
  year = 2024,
  journal = {Phys. Rev. C},
  volume = {109},
  pages = {034302},
  publisher = {American Physical Society},
  url = {https://link.aps.org/doi/10.1103/PhysRevC.109.034302}
}

@article{hohenberg1964,
  title = {Inhomogeneous {{Electron Gas}}},
  author = {Hohenberg, P. and Kohn, W.},
  year = 1964,
  journal = {Phys. Rev.},
  volume = {136},
  pages = {B864-B871},
  issn = {0031-899X},
  url = {https://link.aps.org/doi/10.1103/PhysRev.136.B864},
  copyright = {http://link.aps.org/licenses/aps-default-license}
}

@article{jiang2024,
  title = {Role of Magnetic Fields on the Outer Crust in a Magnetar*},
  author = {Jiang, Wei and Chen, Yan-jun},
  year = 2024,
  journal = {Chinese Phys. C},
  volume = {48},
  pages = {074103},
  publisher = {{Chinese Physical Society and the Institute of High Energy Physics of the Chinese Academy of Sciences and the Institute of Modern Physics of the Chinese Academy of Sciences and IOP Publishing Ltd}},
  issn = {1674-1137},
  url = {https://doi.org/10.1088/1674-1137/ad39cc}
}

@article{jin2017,
  title = {Coordinate-Space Solver for Superfluid Many-Fermion Systems with the Shifted Conjugate-Orthogonal Conjugate-Gradient Method},
  author = {Jin, Shi and Bulgac, Aurel and Roche, Kenneth and Wlaz{\l}owski, Gabriel},
  year = 2017,
  journal = {Phys. Rev. C},
  volume = {95},
  pages = {044302},
  publisher = {American Physical Society},
  url = {https://link.aps.org/doi/10.1103/PhysRevC.95.044302}
}

@article{jin2021,
  title = {The {{LISE}} Package: {{Solvers}} for Static and Time-Dependent Superfluid Local Density Approximation Equations in Three Dimensions},
  shorttitle = {The {{LISE}} Package},
  author = {Jin, Shi and Roche, Kenneth J. and Stetcu, Ionel and Abdurrahman, Ibrahim and Bulgac, Aurel},
  year = 2021,
  journal = {Computer Physics Communications},
  volume = {269},
  pages = {108130},
  issn = {0010-4655},
  url = {https://www.sciencedirect.com/science/article/pii/S0010465521002423}
}

@article{kashiwaba2019,
  title = {Self-Consistent Band Calculation of the Slab Phase in the Neutron-Star Crust},
  author = {Kashiwaba, Yu and Nakatsukasa, Takashi},
  year = 2019,
  journal = {Phys. Rev. C},
  volume = {100},
  pages = {035804},
  publisher = {American Physical Society},
  url = {https://link.aps.org/doi/10.1103/PhysRevC.100.035804}
}

@article{kashiwaba2020,
  title = {Coordinate-Space Solver for Finite-Temperature {{Hartree-Fock-Bogoliubov}} Calculations Using the Shifted {{Krylov}} Method},
  author = {Kashiwaba, Yu and Nakatsukasa, Takashi},
  year = 2020,
  journal = {Phys. Rev. C},
  volume = {101},
  pages = {045804},
  publisher = {American Physical Society},
  url = {https://link.aps.org/doi/10.1103/PhysRevC.101.045804}
}

@article{kaspi2017,
  title = {Magnetars},
  author = {Kaspi, Victoria M. and Beloborodov, Andrei M.},
  year = 2017,
  journal = {Annu. Rev. Astron. Astrophys.},
  volume = {55},
  pages = {261--301},
  publisher = {Annual Reviews},
  issn = {0066-4146, 1545-4282},
  url = {https://www.annualreviews.org/content/journals/10.1146/annurev-astro-081915-023329}
}

@article{kobayashi2023,
  title = {Proximity Effects of Vortices in Neutron \$\textbraceleft\textbraceright\textasciicircum\textbraceleft 3\textbraceright\textbraceleft{{P}}\textbraceright\_\textbraceleft 2\textbraceright\$ Superfluids in Neutron Stars: {{Vortex}} Core Transitions and Covalent Bonding of Vortex Molecules},
  shorttitle = {Proximity Effects of Vortices in Neutron \$\textbraceleft\textbraceright\textasciicircum\textbraceleft 3\textbraceright\textbraceleft{{P}}\textbraceright\_\textbraceleft 2\textbraceright\$ Superfluids in Neutron Stars},
  author = {Kobayashi, Michikazu and Nitta, Muneto},
  year = 2023,
  journal = {Phys. Rev. C},
  volume = {107},
  pages = {045801},
  publisher = {American Physical Society},
  url = {https://link.aps.org/doi/10.1103/PhysRevC.107.045801}
}

@article{kobyakov2013,
  title = {Dynamics of the Inner Crust of Neutron Stars: {{Hydrodynamics}}, Elasticity, and Collective Modes},
  shorttitle = {Dynamics of the Inner Crust of Neutron Stars},
  author = {Kobyakov, D. and Pethick, C. J.},
  year = 2013,
  journal = {Phys. Rev. C},
  volume = {87},
  pages = {055803},
  publisher = {American Physical Society},
  url = {https://link.aps.org/doi/10.1103/PhysRevC.87.055803}
}

@article{kohn1965,
  title = {Self-{{Consistent Equations Including Exchange}} and {{Correlation Effects}}},
  author = {Kohn, W. and Sham, L. J.},
  year = 1965,
  journal = {Phys. Rev.},
  volume = {140},
  pages = {A1133-A1138},
  publisher = {American Physical Society},
  url = {https://link.aps.org/doi/10.1103/PhysRev.140.A1133}
}

@article{lattimer2021,
  title = {Neutron {{Stars}} and the {{Nuclear Matter Equation}} of {{State}}},
  author = {Lattimer, J. M.},
  year = 2021,
  journal = {Annu. Rev. Nucl. Part. Sci.},
  volume = {71},
  url = {https://par.nsf.gov/biblio/10330279-neutron-stars-nuclear-matter-equation-state}
}

@article{leinson2020,
  title = {Vortex Lattice in Rotating Neutron Spin-Triplet Superfluid},
  author = {Leinson, Lev B},
  year = 2020,
  journal = {Mon Not R Astron Soc},
  volume = {498},
  pages = {304--309},
  issn = {0035-8711},
  url = {https://doi.org/10.1093/mnras/staa2475}
}

@article{lesinski2007,
  title = {Tensor Part of the {{Skyrme}} Energy Density Functional: {{Spherical}} Nuclei},
  shorttitle = {Tensor Part of the {{Skyrme}} Energy Density Functional},
  author = {Lesinski, T. and Bender, M. and Bennaceur, K. and Duguet, T. and Meyer, J.},
  year = 2007,
  journal = {Phys. Rev. C},
  volume = {76},
  pages = {014312},
  publisher = {American Physical Society},
  url = {https://link.aps.org/doi/10.1103/PhysRevC.76.014312}
}

@article{link1999,
  title = {Pulsar {{Constraints}} on {{Neutron Star Structure}} and {{Equation}} of {{State}}},
  author = {Link, Bennett and Epstein, Richard I. and Lattimer, James M.},
  year = 1999,
  journal = {Phys. Rev. Lett.},
  volume = {83},
  pages = {3362--3365},
  publisher = {American Physical Society},
  url = {https://link.aps.org/doi/10.1103/PhysRevLett.83.3362}
}

@incollection{lombardo2001,
  title = {Superfluidity in {{Neutron Star Matter}}},
  booktitle = {Physics of {{Neutron Star Interiors}}},
  author = {Lombardo, Umberto and Schulze, Hans-Josef},
  year = 2001,
  volume = {578},
  pages = {30},
  url = {https://ui.adsabs.harvard.edu/abs/2001LNP...578...30L}
}

@article{magierski2002a,
  title = {Structure of the Inner Crust of Neutron Stars: {{Crystal}} Lattice or Disordered Phase?},
  shorttitle = {Structure of the Inner Crust of Neutron Stars},
  author = {Magierski, P. and Heenen, P.-H.},
  year = 2002,
  journal = {Phys. Rev. C},
  volume = {65},
  pages = {045804},
  publisher = {American Physical Society},
  url = {https://link.aps.org/doi/10.1103/PhysRevC.65.045804}
}

@article{makishima2014,
  title = {Possible {{Evidence}} for {{Free Precession}} of a {{Strongly Magnetized Neutron Star}} in the {{Magnetar 4U}} \$0142+61\$},
  author = {Makishima, K. and Enoto, T. and Hiraga, J. S. and Nakano, T. and Nakazawa, K. and Sakurai, S. and Sasano, M. and Murakami, H.},
  year = 2014,
  journal = {Phys. Rev. Lett.},
  volume = {112},
  pages = {171102},
  publisher = {American Physical Society},
  url = {https://link.aps.org/doi/10.1103/PhysRevLett.112.171102}
}

@article{marmorini2024,
  title = {Pulsar Glitches from Quantum Vortex Networks},
  author = {Marmorini, Giacomo and Yasui, Shigehiro and Nitta, Muneto},
  year = 2024,
  journal = {Sci Rep},
  volume = {14},
  pages = {7857},
  publisher = {Nature Publishing Group},
  issn = {2045-2322},
  url = {https://www.nature.com/articles/s41598-024-56383-w},
  copyright = {2024 The Author(s)}
}

@article{maruhn2014,
  title = {The {{TDHF}} Code {{Sky3D}}},
  author = {Maruhn, J. A. and Reinhard, P. -G. and Stevenson, P. D. and Umar, A. S.},
  year = 2014,
  journal = {Computer Physics Communications},
  volume = {185},
  pages = {2195--2216},
  issn = {0010-4655},
  url = {https://www.sciencedirect.com/science/article/pii/S0010465514001313}
}

@article{masaki2020,
  title = {Microscopic Description of Axisymmetric Vortices in \$\textasciicircum\textbraceleft 3\textbraceright{{P}}\_\textbraceleft 2\textbraceright\$ Superfluids},
  author = {Masaki, Yusuke and Mizushima, Takeshi and Nitta, Muneto},
  year = 2020,
  journal = {Phys. Rev. Res.},
  volume = {2},
  pages = {013193},
  publisher = {American Physical Society},
  url = {https://link.aps.org/doi/10.1103/PhysRevResearch.2.013193}
}

@article{masuda2016,
  title = {Magnetic Properties of Quantized Vortices in Neutron \$\textasciicircum\textbraceleft 3\textbraceright{{P}}\_\textbraceleft 2\textbraceright\$ Superfluids in Neutron Stars},
  author = {Masuda, Kota and Nitta, Muneto},
  year = 2016,
  journal = {Phys. Rev. C},
  volume = {93},
  pages = {035804},
  publisher = {American Physical Society},
  url = {https://link.aps.org/doi/10.1103/PhysRevC.93.035804}
}

@article{masuda2020,
  title = {Half-Quantized Non-{{Abelian}} Vortices in Neutron {{3P2}} Superfluids inside Magnetars},
  author = {Masuda, Kota and Nitta, Muneto},
  year = 2020,
  journal = {Prog Theor Exp Phys},
  volume = {2020},
  pages = {013D01},
  issn = {2050-3911},
  url = {https://doi.org/10.1093/ptep/ptz138}
}

@article{minami2022,
  title = {Effects of Pairing Gap and Band Gap on Superfluid Density in the Inner Crust of Neutron Stars},
  author = {Minami, Yuki and Watanabe, Gentaro},
  year = 2022,
  journal = {Phys. Rev. Res.},
  volume = {4},
  pages = {033141},
  publisher = {American Physical Society},
  url = {https://link.aps.org/doi/10.1103/PhysRevResearch.4.033141}
}

@article{muzikar1980,
  title = {\$\textasciicircum\textbraceleft 3\textbraceright{{P}}\_\textbraceleft 2\textbraceright\$ Pairing in Neutron-Star Matter: {{Magnetic}} Field Effects and Vortices},
  shorttitle = {\$\textasciicircum\textbraceleft 3\textbraceright{{P}}\_\textbraceleft 2\textbraceright\$ Pairing in Neutron-Star Matter},
  author = {Muzikar, Paul and Sauls, J. A. and Serene, J. W.},
  year = 1980,
  journal = {Phys. Rev. D},
  volume = {21},
  pages = {1494--1502},
  publisher = {American Physical Society},
  url = {https://link.aps.org/doi/10.1103/PhysRevD.21.1494}
}

@article{nakatsukasa2016,
  title = {Time-Dependent Density-Functional Description of Nuclear Dynamics},
  author = {Nakatsukasa, Takashi and Matsuyanagi, Kenichi and Matsuo, Masayuki and Yabana, Kazuhiro},
  year = 2016,
  journal = {Rev. Mod. Phys.},
  volume = {88},
  pages = {045004},
  publisher = {American Physical Society},
  url = {https://link.aps.org/doi/10.1103/RevModPhys.88.045004}
}

@misc{nam2025,
  title = {Data-Driven Exploration of the Neutron \$\textasciicircum 3\textbackslash text\textbraceleft{{P}}\textbraceright\_2\$ Pairing Gap Using {{Cassiopeia A}} Neutron Star Observational Data},
  author = {Nam, Yoonhak and Sekizawa, Kazuyuki},
  year = 2025,
  eprint = {2510.20353},
  primaryclass = {nucl-th},
  publisher = {arXiv},
  url = {http://arxiv.org/abs/2510.20353},
  archiveprefix = {arXiv}
}

@article{naso2008,
  title = {Magnetic Field Amplification in Proto-Neutron Stars - {{The}} Role of the Neutron-Finger Instability for Dynamo Excitation},
  author = {Naso, L. and Rezzolla, L. and Bonanno, A. and Patern{\`o}, L.},
  year = 2008,
  journal = {A\&A},
  volume = {479},
  pages = {167--176},
  publisher = {EDP Sciences},
  issn = {0004-6361, 1432-0746},
  url = {https://www.aanda.org/articles/aa/abs/2008/07/aa8360-07/aa8360-07.html},
  copyright = {\copyright{} ESO, 2008}
}

@article{newton2009a,
  title = {Modeling Nuclear ``pasta'' and the Transition to Uniform Nuclear Matter with the {{3D Skyrme-Hartree-Fock}} Method at Finite Temperature: {{Core-collapse}} Supernovae},
  shorttitle = {Modeling Nuclear ``pasta'' and the Transition to Uniform Nuclear Matter with the {{3D Skyrme-Hartree-Fock}} Method at Finite Temperature},
  author = {Newton, W. G. and Stone, J. R.},
  year = 2009,
  journal = {Phys. Rev. C},
  volume = {79},
  pages = {055801},
  publisher = {American Physical Society},
  url = {https://link.aps.org/doi/10.1103/PhysRevC.79.055801}
}

@article{ozel2016,
  title = {Masses, {{Radii}}, and the {{Equation}} of {{State}} of {{Neutron Stars}}},
  author = {{\"O}zel, Feryal and Freire, Paulo},
  year = 2016,
  journal = {Annu. Rev. Astron. Astrophys.},
  volume = {54},
  pages = {401--440},
  publisher = {Annual Reviews},
  issn = {0066-4146, 1545-4282},
  url = {https://www.annualreviews.org/content/journals/10.1146/annurev-astro-081915-023322;jsessionid=ZWwwLTVfcj1U9keKMT0RxJUXSKzHwMPdI1sTwPAS.annurevlive-10-241-10-87}
}

@article{page2004,
  title = {Minimal {{Cooling}} of {{Neutron Stars}}: {{A New Paradigm}}},
  shorttitle = {Minimal {{Cooling}} of {{Neutron Stars}}},
  author = {Page, Dany and Lattimer, James M. and Prakash, Madappa and Steiner, Andrew W.},
  year = 2004,
  journal = {Astrophys. J. Suppl. Ser.},
  volume = {155},
  pages = {623--650},
  publisher = {IOP},
  issn = {0067-0049},
  url = {https://ui.adsabs.harvard.edu/abs/2004ApJS..155..623P}
}

@article{page2009,
  title = {{{NEUTRINO EMISSION FROM COOPER PAIRS AND MINIMAL COOLING OF NEUTRON STARS}}},
  author = {Page, Dany and Lattimer, James M. and Prakash, Madappa and Steiner, Andrew W.},
  year = 2009,
  journal = {ApJ},
  volume = {707},
  pages = {1131},
  publisher = {The American Astronomical Society},
  issn = {0004-637X},
  url = {https://doi.org/10.1088/0004-637X/707/2/1131}
}

@article{page2011,
  title = {Rapid {{Cooling}} of the {{Neutron Star}} in {{Cassiopeia A Triggered}} by {{Neutron Superfluidity}} in {{Dense Matter}}},
  author = {Page, Dany and Prakash, Madappa and Lattimer, James M. and Steiner, Andrew W.},
  year = 2011,
  journal = {Phys. Rev. Lett.},
  volume = {106},
  pages = {081101},
  publisher = {American Physical Society},
  url = {https://link.aps.org/doi/10.1103/PhysRevLett.106.081101}
}

@article{pais2012a,
  title = {Exploring the {{Nuclear Pasta Phase}} in {{Core-Collapse Supernova Matter}}},
  author = {Pais, Helena and Stone, Jirina R.},
  year = 2012,
  journal = {Phys. Rev. Lett.},
  volume = {109},
  pages = {151101},
  publisher = {American Physical Society},
  url = {https://link.aps.org/doi/10.1103/PhysRevLett.109.151101}
}

@article{parmar2023,
  title = {Magnetized Neutron Star Crust within Effective Relativistic Mean-Field Model},
  author = {Parmar, Vishal and Das, H. C. and Sharma, M. K. and Patra, S. K.},
  year = 2023,
  journal = {Phys. Rev. D},
  volume = {107},
  pages = {043022},
  publisher = {American Physical Society},
  url = {https://link.aps.org/doi/10.1103/PhysRevD.107.043022}
}

@article{pecak2024,
  title = {Time-{{Dependent Nuclear Energy-Density Functional Theory Toolkit}} for {{Neutron Star Crust}}: {{Dynamics}} of a {{Nucleus}} in a {{Neutron Superfluid}}},
  shorttitle = {Time-{{Dependent Nuclear Energy-Density Functional Theory Toolkit}} for {{Neutron Star Crust}}},
  author = {P{\k e}cak, Daniel and Zdanowicz, Agata and Chamel, Nicolas and Magierski, Piotr and Wlaz{\l}owski, Gabriel},
  year = 2024,
  journal = {Phys. Rev. X},
  volume = {14},
  pages = {041054},
  publisher = {American Physical Society},
  url = {https://link.aps.org/doi/10.1103/PhysRevX.14.041054}
}

@article{penaarteaga2011,
  title = {Nuclear Structure in Strong Magnetic Fields: {{Nuclei}} in the Crust of a Magnetar},
  shorttitle = {Nuclear Structure in Strong Magnetic Fields},
  author = {Pe{\~n}a Arteaga, D. and Grasso, M. and Khan, E. and Ring, P.},
  year = 2011,
  journal = {Phys. Rev. C},
  volume = {84},
  pages = {045806},
  publisher = {American Physical Society},
  url = {https://link.aps.org/doi/10.1103/PhysRevC.84.045806}
}

@article{peng2006,
  title = {The {{Origin}} of {{Glitches}} in {{Pulsars}} --- {{Phase Oscillation}} between {{Anisotropic Superfluid}} and {{Normal State}} of {{Neutrons}} in {{Neutron Stars}}},
  author = {Peng, Qiu-He and Luo, Zhi-Quan and Chou, Chih-Kang},
  year = 2006,
  journal = {Chin. J. Astron. Astrophys.},
  volume = {6},
  pages = {297},
  issn = {1009-9271},
  url = {https://doi.org/10.1088/1009-9271/6/3/04}
}

@misc{potekhin1996a,
  title = {Electron Conduction along Quantizing Magnetic Fields in Neutron Star Crusts. {{II}}. {{Practical}} Formulae.},
  author = {Potekhin, A. Y. and Yakovlev, D. G.},
  year = 1996,
  volume = {314},
  publisher = {arXiv},
  issn = {0004-6361},
  url = {https://ui.adsabs.harvard.edu/abs/1996A&A...314..341P}
}

@article{potekhin2015,
  title = {Neutron {{Stars}}---{{Cooling}} and {{Transport}}},
  author = {Potekhin, Alexander Y. and Pons, Jos{\'e} A. and Page, Dany},
  year = 2015,
  journal = {Space Sci Rev},
  volume = {191},
  pages = {239--291},
  issn = {1572-9672},
  url = {https://doi.org/10.1007/s11214-015-0180-9}
}

@article{ravenhall1983,
  title = {Structure of {{Matter}} below {{Nuclear Saturation Density}}},
  author = {Ravenhall, D. G. and Pethick, C. J. and Wilson, J. R.},
  year = 1983,
  journal = {Phys. Rev. Lett.},
  volume = {50},
  pages = {2066--2069},
  publisher = {American Physical Society},
  url = {https://link.aps.org/doi/10.1103/PhysRevLett.50.2066}
}

@book{ring2004,
  title = {The {{Nuclear Many-Body Problem}}},
  author = {Ring, Peter and Schuck, Peter},
  year = 2004,
  publisher = {Springer Science \& Business Media},
  isbn = {978-3-540-21206-5}
}

@article{schuetrumpf2015,
  title = {Twist-Averaged Boundary Conditions for Nuclear Pasta {{Hartree-Fock}} Calculations},
  author = {Schuetrumpf, B. and Nazarewicz, W.},
  year = 2015,
  journal = {Phys. Rev. C},
  volume = {92},
  pages = {045806},
  publisher = {American Physical Society},
  url = {https://link.aps.org/doi/10.1103/PhysRevC.92.045806}
}

@article{schuetrumpf2019,
  title = {Survey of Nuclear Pasta in the Intermediate-Density Regime: {{Shapes}} and Energies},
  shorttitle = {Survey of Nuclear Pasta in the Intermediate-Density Regime},
  author = {Schuetrumpf, B. and {Mart{\'i}nez-Pinedo}, G. and Afibuzzaman, {\relax Md}. and Aktulga, H. M.},
  year = 2019,
  journal = {Phys. Rev. C},
  volume = {100},
  pages = {045806},
  publisher = {American Physical Society},
  url = {https://link.aps.org/doi/10.1103/PhysRevC.100.045806}
}

@article{schwenk2004,
  title = {Polarization {{Contributions}} to the {{Spin Dependence}} of the {{Effective Interaction}} in {{Neutron Matter}}},
  author = {Schwenk, Achim and Friman, Bengt},
  year = 2004,
  journal = {Phys. Rev. Lett.},
  volume = {92},
  pages = {082501},
  publisher = {American Physical Society},
  url = {https://link.aps.org/doi/10.1103/PhysRevLett.92.082501}
}

@article{sedrakian2019,
  title = {Superfluidity in Nuclear Systems and Neutron Stars},
  author = {Sedrakian, Armen and Clark, John W.},
  year = 2019,
  journal = {Eur. Phys. J. A},
  volume = {55},
  pages = {167},
  issn = {1434-601X},
  url = {https://doi.org/10.1140/epja/i2019-12863-6}
}

@article{sedrakian2025,
  title = {Josephson Currents in Neutron Stars},
  author = {Sedrakian, Armen and Rau, Peter B.},
  year = 2025,
  journal = {Phys. Rev. D},
  volume = {111},
  pages = {023044},
  publisher = {American Physical Society},
  url = {https://link.aps.org/doi/10.1103/PhysRevD.111.023044}
}

@article{sekizawa2013,
  title = {Time-Dependent {{Hartree-Fock}} Calculations for Multinucleon Transfer Processes in \$\textbraceleft\textbraceright\textasciicircum\textbraceleft 40,48\textbraceright\${{Ca}}+\$\textbraceleft\textbraceright\textasciicircum\textbraceleft 124\textbraceright\${{Sn}}, \$\textbraceleft\textbraceright\textasciicircum\textbraceleft 40\textbraceright\${{Ca}}+\$\textbraceleft\textbraceright\textasciicircum\textbraceleft 208\textbraceright\${{Pb}}, and \$\textbraceleft\textbraceright\textasciicircum\textbraceleft 58\textbraceright\${{Ni}}+\$\textbraceleft\textbraceright\textasciicircum\textbraceleft 208\textbraceright\${{Pb}} Reactions},
  author = {Sekizawa, Kazuyuki and Yabana, Kazuhiro},
  year = 2013,
  journal = {Phys. Rev. C},
  volume = {88},
  pages = {014614},
  publisher = {American Physical Society},
  url = {https://link.aps.org/doi/10.1103/PhysRevC.88.014614}
}

@article{sekizawa2022,
  title = {Time-Dependent Extension of the Self-Consistent Band Theory for Neutron Star Matter: {{Anti-entrainment}} Effects in the Slab Phase},
  shorttitle = {Time-Dependent Extension of the Self-Consistent Band Theory for Neutron Star Matter},
  author = {Sekizawa, Kazuyuki and Kobayashi, Sorataka and Matsuo, Masayuki},
  year = 2022,
  journal = {Phys. Rev. C},
  volume = {105},
  pages = {045807},
  publisher = {American Physical Society},
  url = {https://link.aps.org/doi/10.1103/PhysRevC.105.045807}
}

@misc{sekizawa2023,
  title = {Possible {{Existence}} of {{Extremely Neutron-Rich Superheavy Nuclei}} in {{Neutron Star Crusts Under}} a {{Superstrong Magnetic Field}}},
  author = {Sekizawa, Kazuyuki and Kaba, Kentaro},
  year = 2023,
  eprint = {2302.07923},
  primaryclass = {nucl-th},
  publisher = {arXiv},
  url = {http://arxiv.org/abs/2302.07923},
  archiveprefix = {arXiv}
}

@article{shternin2011,
  title = {Cooling Neutron Star in the {{Cassiopeia A}} Supernova Remnant: Evidence for Superfluidity in the Core},
  shorttitle = {Cooling Neutron Star in the {{Cassiopeia A}} Supernova Remnant},
  author = {Shternin, Peter S. and Yakovlev, Dmitry G. and Heinke, Craig O. and Ho, Wynn C. G. and Patnaude, Daniel J.},
  year = 2011,
  journal = {Mon. Not. R. Astron. Soc.},
  volume = {412},
  pages = {L108-L112},
  publisher = {OUP},
  issn = {0035-8711},
  url = {https://ui.adsabs.harvard.edu/abs/2011MNRAS.412L.108S}
}

@article{sigrist1991,
  title = {Phenomenological Theory of Unconventional Superconductivity},
  author = {Sigrist, Manfred and Ueda, Kazuo},
  year = 1991,
  journal = {Rev. Mod. Phys.},
  volume = {63},
  pages = {239--311},
  publisher = {American Physical Society},
  url = {https://link.aps.org/doi/10.1103/RevModPhys.63.239}
}

@article{sigrist2005,
  title = {Introduction to {{Unconventional Superconductivity}}},
  author = {Sigrist, Manfred},
  year = 2005,
  journal = {AIP Conf. Proc.},
  volume = {789},
  pages = {165--243},
  issn = {0094-243X},
  url = {https://doi.org/10.1063/1.2080350}
}

@article{stein2016,
  title = {Carbon-Oxygen-Neon Mass Nuclei in Superstrong Magnetic Fields},
  author = {Stein, Martin and Maruhn, Joachim and Sedrakian, Armen and Reinhard, P.-G.},
  year = 2016,
  journal = {Phys. Rev. C},
  volume = {94},
  pages = {035802},
  publisher = {American Physical Society},
  url = {https://link.aps.org/doi/10.1103/PhysRevC.94.035802}
}

@article{stein2016a,
  title = {Spin-Polarized Neutron Matter: {{Critical}} Unpairing and {{BCS-BEC}} Precursor},
  shorttitle = {Spin-Polarized Neutron Matter},
  author = {Stein, Martin and Sedrakian, Armen and Huang, Xu-Guang and Clark, John W.},
  year = 2016,
  journal = {Phys. Rev. C},
  volume = {93},
  pages = {015802},
  issn = {2469-9985, 2469-9993},
  url = {https://link.aps.org/doi/10.1103/PhysRevC.93.015802},
  copyright = {http://link.aps.org/licenses/aps-default-license}
}

@article{tajima2023,
  title = {Exploring \$\textasciicircum\textbraceleft 3\textbraceright{{P}}\_\textbraceleft 0\textbraceright\$ Superfluid in Dilute Spin-Polarized Neutron Matter},
  author = {Tajima, Hiroyuki and Funaki, Hiroshi and Sekino, Yuta and Yasutake, Nobutoshi and Matsuo, Mamoru},
  year = 2023,
  journal = {Phys. Rev. C},
  volume = {108},
  pages = {L052802},
  publisher = {American Physical Society},
  url = {https://link.aps.org/doi/10.1103/PhysRevC.108.L052802}
}

@article{turolla2015,
  title = {Magnetars: The Physics behind Observations. {{A}} Review},
  shorttitle = {Magnetars},
  author = {Turolla, R and Zane, S and Watts, A L},
  year = 2015,
  journal = {Rep. Prog. Phys.},
  volume = {78},
  pages = {116901},
  publisher = {IOP Publishing},
  issn = {0034-4885},
  url = {https://doi.org/10.1088/0034-4885/78/11/116901}
}

@article{watanabe2017,
  title = {Superfluid {{Density}} of {{Neutrons}} in the {{Inner Crust}} of {{Neutron Stars}}: {{New Life}} for {{Pulsar Glitch Models}}},
  shorttitle = {Superfluid {{Density}} of {{Neutrons}} in the {{Inner Crust}} of {{Neutron Stars}}},
  author = {Watanabe, Gentaro and Pethick, C. J.},
  year = 2017,
  journal = {Phys. Rev. Lett.},
  volume = {119},
  pages = {062701},
  publisher = {American Physical Society},
  url = {https://link.aps.org/doi/10.1103/PhysRevLett.119.062701}
}

@article{wei2020,
  title = {Nuclear {{Pairing Gaps}} and {{Neutron Star Cooling}}},
  author = {Wei, Jin-Biao and Burgio, Fiorella and Schulze, Hans-Josef},
  year = 2020,
  journal = {Universe},
  volume = {6},
  pages = {115},
  publisher = {Multidisciplinary Digital Publishing Institute},
  issn = {2218-1997},
  url = {https://www.mdpi.com/2218-1997/6/8/115},
  copyright = {http://creativecommons.org/licenses/by/3.0/}
}

@article{yakovlev2004,
  title = {Neutron {{Star Cooling}}},
  author = {Yakovlev, D. G. and Pethick, C. J.},
  year = 2004,
  journal = {Annu. Rev. Astron. Astrophys.},
  volume = {42},
  pages = {169--210},
  publisher = {Annual Reviews},
  issn = {0066-4146, 1545-4282},
  url = {https://www.annualreviews.org/content/journals/10.1146/annurev.astro.42.053102.134013}
}

@article{yasui2019,
  title = {Phase Structure of Neutron \$\textasciicircum\textbraceleft 3\textbraceright{{P}}\_\textbraceleft 2\textbraceright\$ Superfluids in Strong Magnetic Fields in Neutron Stars},
  author = {Yasui, Shigehiro and Chatterjee, Chandrasekhar and Nitta, Muneto},
  year = 2019,
  journal = {Phys. Rev. C},
  volume = {99},
  pages = {035213},
  publisher = {American Physical Society},
  url = {https://link.aps.org/doi/10.1103/PhysRevC.99.035213}
}

@article{yoshimura2024a,
  title = {Superfluid Extension of the Self-Consistent Time-Dependent Band Theory for Neutron Star Matter: {{Anti-entrainment}} versus Superfluid Effects in the Slab Phase},
  shorttitle = {Superfluid Extension of the Self-Consistent Time-Dependent Band Theory for Neutron Star Matter},
  author = {Yoshimura, Kenta and Sekizawa, Kazuyuki},
  year = 2024,
  journal = {Phys. Rev. C},
  volume = {109},
  pages = {065804},
  publisher = {American Physical Society},
  url = {https://link.aps.org/doi/10.1103/PhysRevC.109.065804}
}

@article{yoshimura2025,
  title = {Phase Transitions in the Inner Crust of Neutron Stars within the Superfluid Band Theory: {{Competition}} between \$\textasciicircum\textbraceleft 1\textbraceright{{S}}\_\textbraceleft 0\textbraceright\$ Pairing and Spin Polarization under Finite Temperature and Magnetic Field},
  shorttitle = {Phase Transitions in the Inner Crust of Neutron Stars within the Superfluid Band Theory},
  author = {Yoshimura, Kenta and Sekizawa, Kazuyuki},
  year = 2025,
  journal = {Phys. Rev. C},
  volume = {112},
  pages = {065804},
  publisher = {American Physical Society},
  url = {https://link.aps.org/doi/10.1103/9yfb-rfdd}
}

@article{zhou2022,
  title = {Pulsar {{Glitches}}: {{A Review}}},
  shorttitle = {Pulsar {{Glitches}}},
  author = {Zhou, Shiqi and G{\"u}gercino{\u g}lu, Erbil and Yuan, Jianping and Ge, Mingyu and Yu, Cong},
  year = 2022,
  journal = {Universe},
  volume = {8},
  pages = {641},
  publisher = {Multidisciplinary Digital Publishing Institute},
  issn = {2218-1997},
  url = {https://www.mdpi.com/2218-1997/8/12/641},
  copyright = {http://creativecommons.org/licenses/by/3.0/}
}
